\documentclass[11pt]{article}
\usepackage[dvips]{graphicx}
\usepackage{amsmath,amssymb}
\usepackage{color}
\usepackage{amscd}

\baselineskip=\normalbaselineskip
\renewcommand{\baselinestretch}{1.4}
\makeatletter
\@addtoreset{equation}{section}

\makeatother
\newcommand{\be}{\begin{equation}}
\newcommand{\ee}{\end{equation}}
\newcommand{\ba}{\begin{eqnarray}}
\newcommand{\ea}{\end{eqnarray}}
\newcommand{\nn}{\nonumber}

\newcommand{\del}{\partial}
\newcommand{\bra}[1]{\left\langle\,{#1}\,\right|}
\newcommand{\ket}[1]{\left|\,{#1}\,\right\rangle}

\newcommand{\bX}{{\mathbf X}}
\newcommand{\bZ}{{\mathbf Z}}

\newcommand{\bP}{\mathbf P}
\newcommand{\bM}{\mathbf M}

\newcommand{\hP}{\widehat{P}}

\newcommand{\hsigma}{\widehat{\sigma}}

\newcommand{\bA}{\mathbf A}

\newcommand{\cP}{{\cal P}}

\newcommand{\Tr}{{\rm Tr\,}}
\newcommand{\Str}{{\rm Str\,}}
\newcommand{\Z}{{\mathbb Z}}
\newcommand{\R}{{\mathbb R}}

\begin{document}
\setcounter{page}{0}

\begin{flushright}
\parbox{40mm}{%
RIKEN-TH-69 \\
hep-th/0604104 \\
April 2006}
\end{flushright}

\vfill

\begin{center}
{\Large{\bf 
Construction of Instantons via Tachyon Condensation
}}
\end{center}

\vfill

\renewcommand{\baselinestretch}{1.0}

\begin{center}
\textsc{Tsuguhiko Asakawa}
\footnote{E-mail: \texttt{t.asakawa@riken.jp}}, 
\textsc{So Matsuura}
\footnote{E-mail: \texttt{matsuso@riken.jp}} and
\textsc{Kazutoshi Ohta}
\footnote{E-mail: \texttt{k-ohta@riken.jp}}

\textsl{Theoretical Physics Laboratory, \\ 
      The Institute of Physical and Chemical Research (RIKEN), \\
     2-1 Hirosawa, Wako, Saitama 351-0198, JAPAN } \\ 
\end{center}

\begin{center}
{\bf abstract}
\end{center}

\begin{quote}

\small{%
We investigate the D-brane bound states 
from the viewpoint of the unstable $D/\overline{D}$-system and their 
tachyon condensation.
We consider two systems; 
a system of $k D(-1)$-branes and $N D3$-branes with 
open strings connecting them and a system of $N D3$-branes with
open strings corresponding to the $k$-instanton flux,  
both of which are realized through the tachyon condensation
from $(N+2k) D3$-branes and $2k \overline{D3}$-branes with
appropriate tachyon profiles.
It can be shown that these systems are related with each other 
through a unitary gauge transformation of the $D3/\overline{D3}$-system.
We construct an explicit form of the gauge transformation and 
show that the essential elements of the ADHM construction
naturally arise from the explicit form of the gauge transformation.
As a result, the ADHM construction is understood as an outcome
of this gauge equivalence in different low energy limits. 
The small instanton singularities can be also understood in this
context. 
Other kinds of solitons with different codimensions are also discussed
from the view point of the tachyon condensation.
}
\end{quote}
\vfill

\renewcommand{\baselinestretch}{1.4}

\renewcommand{\thefootnote}{\arabic{footnote}}
\setcounter{footnote}{0}
\addtocounter{page}{1}
\newpage
\section{Introduction}
\label{Intro}
Solitons play essential roles to understand non-perturbative
aspects of both gauge and string theory.
In gauge theory, solitons are non-trivial solutions for
field equations whose properties often reflects 
the non-perturbative nature of non-abelian gauge theories. 
Solitons are classified by their codimension and  
called as instanton, monopole, vortex and domain wall 
corresponding to the codimensions 4, 3, 2 and 1, respectively,
each of them has its own characteristic properties. 
In particular, solitons in supersymmetric gauge theories
are closely related to D-branes in the superstring theory  
\cite{Douglas:1995bn, Douglas:1996uz,
Diaconescu:1996rk, Witten:1995gx}.
In fact, many BPS solitons are realized as a BPS bound state
of D-branes in the superstring theory.
Once such an identification is given,
we can use many powerful techniques of the superstring theory
to examine the nature of the BPS solitons.

Among solitons in gauge theories, 
we are mainly interested in instantons in this paper.
In general, 
instantons in Yang-Mills theory are classical configurations 
of the gauge field with (anti-)self-dual field strength. 
In constructing the instanton solution, 
the systematic way proposed by
Atiyah, Drinfeld, Hitchin and Manin (ADHM) is quite powerful
\cite{Atiyah:1978ri},
which made us clear to see the structure of instanton moduli space
(see e.g. \cite{Dorey:2002ik} for a review).
Although the ADHM construction is originally developed for 
investigating instantons in non-supersymmetric gauge theories, 
the physical meaning becomes clearer when one consider 
instantons in supersymmetric gauge theories and 
realize them as bound states of D-branes. 
In order to see it, 
let us consider bound states of D-branes with codimension 4, 
such as $D(-1)$-$D3$ bound states, $D0$-$D4$ bound
states, and so on, 
which is a realization of instantons in the ${\cal N}=4$ 
supersymmetric gauge theory.
Although they are equivalent under the T-dualities,
we concretely treat the $D(-1)$-$D3$ bound states hereafter. 
There are two different descriptions of the 
$k D(-1)$-$N D3$ bound states at low energy:
\begin{itemize}
\item[(a)] ADHM data in the $0D$ effective theory on the $k$
	   $D(-1)$-branes
\item[(b)] gauge instanton in the $4D$ effective gauge theory
	   on the $N D3$-branes
\end{itemize}
The latter picture (b) is known as the ``brane within brane''
\cite{Douglas:1995bn, Douglas:1996uz}: 
$k D(-1)$-branes are dissolved into
$N D3$ branes. 
At the low energy limit $\alpha' \rightarrow 0$ with
fixing the gauge coupling on the $D3$-branes,
$g_4^2 \sim g_s$, 
the effective theory on the $N D3$-branes is given by $4D$
${\cal N}=4$ $U(N)$ supersymmetric Yang-Mills theory 
and a self-dual gauge field with the second Chern class $k$ 
induces $k D(-1)$-brane charge through the Chern-Simons coupling 
$C\int \Tr(F\wedge F)$.
On the other hand, in the picture (a),
the low energy limit is taken as $\alpha'\to 0$ with
fixing the gauge coupling on the $D(-1)$-branes,
$g_0^2\sim g_s\alpha'^{-2}$ and the effective theory  
is a $0$-dimensional $U(k)$ supersymmetric ``gauge theory''
on the $D(-1)$-branes 
with hypermultiplets which come from open strings
connecting with $D3$-branes.
The ADHM data is identified with the field configuration
of this $0D$ gauge theory. 
The ADHM construction provides
a remarkable correspondence between (a) and (b).
In fact, 
the D- and F-flatness conditions of (a),
which is the ADHM equations, 
is equivalent to the 
self-duality of the $4D$ gauge field on $D3$-branes. 
Furthermore, 
the moduli space of the Higgs vacua of the $0D$ gauge theory,
which is the hyper-K\"ahler quotient 
of the ADHM equations by $U(k)$ gauge group, 
agrees exactly with the instanton moduli space. 
The Higgs and Coulomb phase in the $0D$ gauge theory
are connected with each other through the small instanton singularity in (b).
In this manner, 
we can understand the properties of the instantons of the $4D$ gauge theory
from the $0D$ gauge theory through the ADHM construction.
However, we emphasize here that
these two descriptions (a) and (b) are 
valid in the very different low energy limit of the
$D(-1)$-$D3$ bound state 
so that the process of the ADHM construction itself 
is still non-trivial
in the D-brane interpretation.

Besides, by considering the instantons in terms of D-branes, 
we can use many powerful techniques of the superstring theory 
to support the correspondence between the two descriptions further.
For example, the properties in the construction of the instanton solution
like the reciprocity \cite{Corrigan:1983sv},
relation to the Nahm construction of monopoles \cite{Nahm:1979yw},
and Fourier-Mukai transformation \cite{Hori:1999me} 
can now be understood as a kind of
T-duality in the superstring theory which reduces or extends
the world-volume directions of D-branes.
These identifications also can be applied to the non-commutative 
spacetime case \cite{Nekrasov:1998ss}.
The non-commutativity resolves the small instanton singularity 
by turning on the Fayet-Iliopoulos parameter to 
the D- and F-flatness conditions.
So the non-commutativity prevents entering the Coulomb phase,
namely the $D(-1)$-$D3$ bound states cannot be separated with each other.
Therefore, the correspondence between (a) and (b) at low energy is still 
expected to hold at the level of the bound state of D-branes 
in the full string theory \cite{Billo:2002hm, Billo:2005fg},
that is, it does not rely on the low energy limit.
To see this, it is convenient to work with the boundary states, 
since they carry all the information on the D-branes, 
independent of $\alpha' \rightarrow 0$ limit.
One of the purpose of this paper is to show the equivalence of two 
descriptions at the level of the boundary states. 
In other words, we will clarify the meaning of the 
ADHM construction in the string theory.

The key observation is that $D(-1)$-branes are realized 
by the tachyon condensation in the unstable $D3$/$\overline{D3}$ system, 
known as the D-brane descent relation \cite{Sen:1998sm}.
This indicates that not only $D(-1)$-branes but also 
the $D(-1)$-$D3$ bound states are 
described by the $D3/\overline{D3}$ system.
In this case, we should consider different number
of $D3$-branes and $\overline{D3}$-branes, where the tachyon becomes
a rectangular matrix and it makes the structure of this system non-trivial\footnote{
See \cite{Jones:2003ae,Ishida:2006tj} for the other discussions 
dealing with different number of D-branes and anti-D-branes.}.
In this sense, it is our another motivation of this paper to establish a
way to investigate the D-brane bound states in terms of
systems with different number of D-branes and
anti-D-branes.
In order to treat such $D/\overline{D}$ systems in the full string
theory, it is convenient to use the boundary state formalism
\cite{Asakawa:2002ui}.

Along this line, there appeared the literature 
\cite{Akhmedov:2001jq} to describe $D5$-branes inside the $D9$-branes
in the context of the low energy effective action for the
$D9/\overline{D9}$ system.   
They have identified a tachyon profile on $D9/\overline{D9}$ system as 
the ADHM data, evaluated the Chern-Simons term in the simplest example 
and confirmed that it possesses the correct RR-charge density
expected from the instanton configuration.
More recently, in the nice paper \cite{Hashimoto:2005qh}, 
not only the ADHM data but also the ADHM construction itself 
are considered in the boundary state formalism for 
the $D4/\overline{D4}$ system.
They began with the same tachyon profile as \cite{Akhmedov:2001jq}, 
which describes a bound state of $D0$-branes and $D4$-branes, 
and considered the gauge transformation in the $D4/\overline{D4}$ system 
with diagonalizing the tachyon profile itself is used.
They showed that 
$D4$-branes and non-trivial gauge field on them are obtained
after the tachyon condensation.
Here the technique to evaluate the boundary state has already developed 
by the same authors in \cite{Hashimoto:2005yy} in the case of the Nahm 
construction of monopoles.

In this paper, we investigate further the D-brane system in more detail, 
which is equivalent to that treated in \cite{Akhmedov:2001jq,Hashimoto:2005qh}.
\begin{figure}[t]
\begin{center}
\includegraphics[scale=0.55]{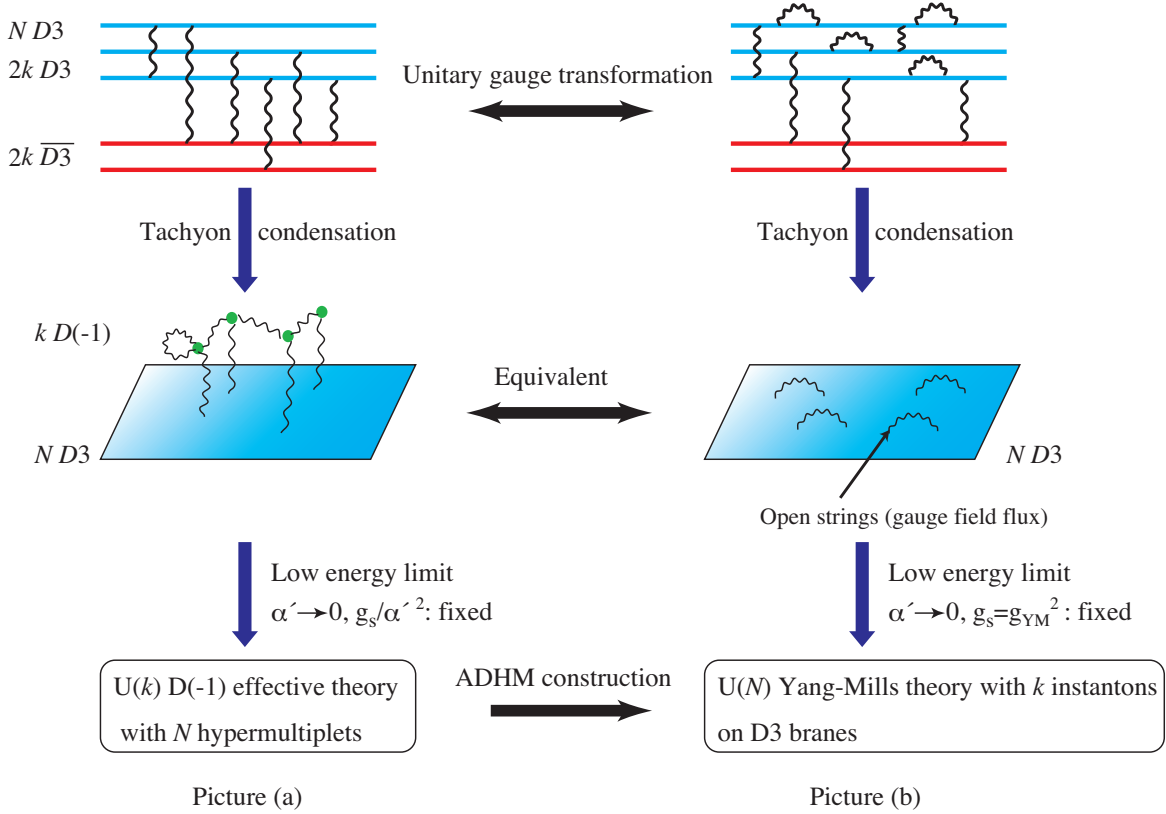}
\end{center}
\caption{{\small
A sketch of the whole structure. 
In the middle, the left describes $D(-1)$-$D3$ bound state 
with DN- and DD-open strings,
while the right one is $D3$-branes with open strings on them.
Both are realized in $D3$/$\overline{D3}$-system and 
are gauge equivalent with each other.
In the low energy limit, each reduces to the
effective theory on the $D(-1)$-branes (picture (a)) 
and the $D3$-branes (picture (b)), respectively. 
The relation at the low energies is the ADHM construction.
}}
\label{strategy}
\end{figure}
Following the basic prescription of \cite{Hashimoto:2005qh},
we pay attention to the difference of roles of the gauge transformation
and the tachyon condensation. 
Here, let us briefly summarize our strategy (see Fig.~\ref{strategy}).
We investigate the tachyon condensation of the system
of $(N+2k) D3$-branes and $2k \overline{D3}$-branes in two ways
corresponding to the pictures (a) and (b).
We first consider the tachyon profile representing 
individual $k D(-1)$-branes and $N D3$-branes and 
add fluctuations of open strings ending on $D(-1)$-branes
(DD-strings and DN-strings).
This is the picture (a) where the system
is regarded as a $k D(-1)$-$N D3$ bound state 
with open strings between them.
Next, we consider a unitary transformation  
in the $D3$/$\overline{D3}$-system 
that shuffles the Chan-Paton indices of $(N+2k) D3$-branes.
We show that new set of $2k D3$/$\overline{D3}$-branes 
completely annihilate into the vacuum under the tachyon condensation 
and instead the $k$-instanton gauge field (NN strings) appears 
on the remaining $N D3$-branes.
This gives the picture (b) where the system is regarded as $N$
$D3$-branes with open strings.
Since the two pictures are gauge equivalent with each other in the
$D3/\overline{D3}$ system, 
they are different realizations of the same system.  
From this point of view, the ADHM construction is understood
as a one-to-one correspondence between different low energy limits
of the pictures (a) and (b). 

Our novel departures in this paper are in the following. 
First, we show that the mechanism to produce a non-trivial gauge flux 
on the remaining $D$-branes presented in \cite{Hashimoto:2005qh} 
is not restricted to the ADHM construction but a general consequence of 
the rectangular tachyon. 
Next, we explicitly present the concrete expression for the 
gauge transformation on the
$D3/\overline{D3}$-brane system 
and take into account the worldsheet fermions not only the bosons,
which are already promised in \cite{Hashimoto:2005yy,Hashimoto:2005qh}.
As a result, we find that the completeness condition of the ADHM
construction appears naturally so that the remaining gauge configuration 
is shown to be self-dual.
It also guarantees the correct RR-charges. 
Moreover, the non-existence of the gauge transformation is shown to 
correspond to the small 
instanton singularity which spoils the ADHM construction of the
self-dual gauge field strength.
Finally, we emphasize the physical
interpretation of the procedure based on the concept of
``picture'' explained above. 
The physical meaning of the ADHM construction becomes even clearer
by introducing this concept.
The relation between boundary states and low energy effective theory, 
the moduli space in terms of boundary states and the treatment of off-shell
boundary states are easily understood.

The above arguments are not only specialized
for the $D(-1)$-$D3$ bound state but also can be applied to
other kind of D-brane bound states with even codimensions. 
Thus we will start from generic discussions on
$Dp$/$\overline{Dp}$ system with $p+1=2n$, 
and after that we will go into the details
of a specific bound state.
In the low energy limit (of the picture (b)), the bound state would
correspond to a BPS soliton if we impose the BPS condition to the
system. In this way, we can apply our method to construct BPS solitons in
principle.

The organization of the paper is as follows.
In the next section,
we briefly give generic arguments
on the tachyon condensation in the system of
the different number of $Dp$-branes and $\overline{Dp}$-branes
in terms of the boundary state 
and review the construction of
$D(-1)$-branes via the tachyon condensation
as a preliminary of the discussion in the following section.
In the section 3, 
we give a general mechanism to produce a non-trivial gauge 
configuration on the remaining $N Dp$-branes
after the tachyon condensation. 
In the section 4,
we apply the procedure of the section 3 to the case of $p=3$.
We express the $D(-1)$-$D3$ bound state in both of the pictures (a) and
(b), and show that they are gauge equivalent with each other in the full
string theory. We see that a gauge field naturally appears
from the gauge transformation and it becomes an instanton gauge field if
we impose the BPS condition.   
We also discuss the small instanton singularities and off-shell 
configurations of the bound state. 
In the section 5, we apply our method to other kinds of D-brane bound
states. We discuss the relation between the bound state and higher
dimensional instantons. We show that the $8$-dimensional version of the
ADHM construction given in \cite{Ohta:2001dh} is reproduced from our
method.

\section{Tachyon Condensation in $Dp$/$\overline{Dp}$ system}

In this section, we would like to discuss general properties 
on the tachyon condensation of the different numbers of the 
$Dp$/$\overline{Dp}$ system in terms of the (off-shell) 
boundary state formalism.
It is a straightforward generalization of \cite{Asakawa:2002ui}, 
and is closely related to the boundary string field theory (BSFT)
\cite{Kraus:2000nj}\cite{Takayanagi:2000rz}.

\subsection{$Dp$/$\overline{Dp}$ system in terms of the boundary state}

Let us first introduce the boundary states for $(N+M) Dp$-branes 
and $M \overline{Dp}$-branes in the Type II superstring theory.  
The boundary state is a state in the closed string Hilbert 
space defined by 
\begin{equation}
 \ket{Bp}_{\rm NS(RR)} \equiv
 \int[d\bX^\mu]\ket{\bX^\mu,\bX^i=0}_{\rm NS(RR)}, 
 \label{Bp}
\end{equation}
where $\mu=0,\cdots,p$ and $i=p+1,\cdots,9$ represent
the worldvolume and the transverse directions
of the $Dp$-branes, respectively.
$\ket{\bX^M}_{\rm NS(RR)}$ $(M=(\mu,i))$
are the eigenstates of the worldsheet closed string superfields 
restricted on the boundary, 
\begin{equation}
\bX^M(\hsigma) = X^M(\sigma) + 
\sqrt{\alpha'}\theta \Psi^M(\sigma), 
\label{boundary superfield}
\end{equation}
in the NSNS(RR)-sector.
In this paper we set $\alpha'=1$. 
In (\ref{boundary superfield}), the superfields are functions in the
boundary super coordinate $\hsigma=(\sigma,\theta)$ unifying the 
worldsheet bosonic coordinate $0\le\sigma < 2\pi$
and its super-partner $\theta$.
We have omitted the ghost contribution and the sign $\pm$ for the spin structure
since they play no role for our purpose.
The boundary state (\ref{Bp})
represents the Neumann boundary condition for the $\mu$ direction 
while the Dirichlet one for the $i$ direction.

The open string degrees of freedom on the D-branes 
are introduced through the 
boundary interaction $S_b$ acting on the boundary state as  
\begin{equation}
 e^{-S_b}\ket{Bp}_{\rm NS(RR)}, 
\end{equation}
where $e^{-S_b}$ has the form, 
\begin{equation}
e^{-S_b} \equiv 
\begin{cases}
\Tr \hP e^{\int d\hsigma \bM(\hsigma)} & \quad\text{(NSNS)}, \\
\Str \hP e^{\int d\hsigma \bM(\hsigma)} & \quad\text{(RR)}, 
\end{cases}
\label{boundary interaction}
\end{equation}
through a matrix $\bM(\hsigma)$.  
$\hP$ in (\ref{boundary interaction}) denotes
the super path-ordered product, 
\begin{equation}
\hP e^{\int d\hsigma \bM(\hsigma)}
\equiv 1+ \sum_{n=1}^\infty (-1)^{\frac{n(n-1)}{2}}
\int d\hsigma_1\cdots d\hsigma_n 
\Theta(\hsigma_1-\hsigma_2)\cdots\Theta(\hsigma_{n-1}-\hsigma_n)
\bM(\hsigma_1)\cdots\bM(\hsigma_n),
\label{super path-ordering}
\end{equation}
with the supersymmetric Heaviside function, 
$\Theta(\hsigma-\hsigma')\equiv\theta(\sigma-\sigma')
-\theta\theta' \delta(\sigma-\sigma')$. 
In the system with $(N+M) Dp$-branes and $M$ 
$\overline{Dp}$-branes, the Chan-Paton Hilbert space is 
$(N+2M)$-dimensional complex vector space,
so that,  
$\bM(\hsigma)$ is an $ (N+2M) \times (N+2M)$ matrix.
Then the trace (\ref{boundary interaction}) in the NSNS-sector 
is taken over the Chan-Paton space ${\mathbb C}^{N+2M}$.
It also possesses a $\Z_2$-grading 
with respect to the sign for the RR-charge.
Then the super-trace in the RR-sector 
of (\ref{boundary interaction}) is defined by
inserting the grading operator $\Gamma$:
\begin{equation}
\Str ( \hP \cdots )
= \Tr 
(\Gamma \hP \cdots), \qquad
\Gamma=\left(
\begin{array}{cc}
{\mathbf 1}_{N+M} & 0 \\
0 & -{\mathbf 1}_{M} 
\end{array}
\right). 
\end{equation}
We sometimes write $\bM(\hsigma)$ in components with respect to $\theta$ as
\begin{equation}
 \bM(\hsigma) = M_0(\sigma) + \theta M_1(\sigma). 
\end{equation}
Using these component matrices, 
we can carry out $\theta$ integral in the definition 
of the boundary interaction (\ref{boundary interaction}),
then it becomes  
\begin{equation}
\hP e^{\int d\hsigma \bM(\hsigma)}
= P e^{\int d\sigma M(\sigma)},
\label{path order}
\end{equation}
with 
\begin{equation}
M(\sigma) \equiv M_1(\sigma) - M_0(\sigma)^2, 
\label{super field stringth}
\end{equation}
where $P$ now stands for the standard path-ordered product.

An important property 
is that this system has $U(N+M)\!\times\! U(M)$ gauge symmetry.
The boundary interaction (\ref{boundary interaction}) is invariant 
under the gauge transformation of the form, 
\begin{equation}
\bM \to \bM'=-U(\bX)^\dagger DU(\bX)
+ U(\bX)^\dagger \bM U(\bX), 
\label{gauge transformation}
\end{equation}
with 
\begin{equation}
U(\bX) = \left(
\begin{array}{cc}
U_1(\bX) & 0 \\
0 & U_2(\bX) 
\end{array}
\right),
\qquad U_1\in U(N+M),\ \ U_2 \in U(M),
\label{gauge function}
\end{equation}
where $D=\del_\theta +  \theta \del_\sigma$ 
is the boundary super-derivative and 
$U_1(U_2)$ is an arbitrary functional of $\bX^\mu$ 
valued in $U(N\!+\!M)(U(M))$, respectively.
The first term in (\ref{gauge transformation}) is necessary for the
invariance  
which comes from the definition of the super path-ordering 
(\ref{super path-ordering}).
It is a super-analog of the gauge invariance of the Wilson loop operator
(for the proof see Appendix \ref{app gauge}).

When we are interested in the massless and tachyonic excitations 
of open strings on this system, $\bM$ can be written as 
\begin{equation}
\bM = \left(
\begin{array}{cc}
A_\mu(\bX^\mu) D\bX^\mu + \Phi^i(\bX^\mu) \bP_i & T(\bX^\mu) \\
T(\bX^\mu)^\dagger & \widetilde{A}_\mu(\bX^\mu) D\bX^\mu
+ \widetilde{\Phi}^i(\bX^\mu) \bP_i
\end{array}
\right), 
\label{general superconnection}
\end{equation}
where $\bP_i$ is the conjugate momentum operator for $\bX^i$. 
It represents boundary insertions of the massless and tachyon
vertex operators with coefficient as corresponding fields on the 
$Dp$-branes.
In our situation, the gauge field $A_\mu$ and scalar fields $\Phi_i$ 
($\widetilde{A}_\mu$ and $\widetilde{\Phi}_i$)
on the $Dp$-branes ($\overline{Dp}$-branes)
are valued in ${(N+M)}\!\times\!{(N+M)}$
(${M}\!\times\!{M}$) matrices, respectively,
which are transformed
in the adjoint representation of the gauge group 
$U(N+M)$ and $U(M)$, respectively. 
Similarly the tachyon field $T$ connecting between 
the $Dp$-branes and the $\overline{Dp}$-branes
is valued in $(N+M)\!\times\!{M}$ complex matrix 
which is transformed in the bi-fundamental representation
$(N+M,\bar{M})$ of $U(N+M)\!\times\! U(M)$.

Since we would like to consider the situation
that the initial massless fluctuation
is absent and the only tachyon field is turned on, 
we simply drop the other fluctuation except for tachyon 
in (\ref{general superconnection}) as 
\begin{equation}
\bM = \left(
\begin{array}{cc}
0 & T(\bX) \\
T(\bX)^\dagger & 0
\end{array}
\right),
\label{tachyon superconnection}
\end{equation}
which also gives rise to $M$ in (\ref{super field stringth}) as
\begin{equation}
M = \left(
\begin{array}{cc}
-TT^\dagger (X) & \Psi^\mu \del_\mu T (X)\\
\Psi^\mu \del_\mu T^\dagger (X) & -T^\dagger T (X)
\end{array}
\right).
\label{tachyon supercurvature}
\end{equation}
By a gauge transformation (\ref{gauge transformation}), 
this term (\ref{tachyon superconnection}) transforms into 
\begin{equation}
\bM'=\left(
\begin{array}{cc}
-U_1^\dagger (\bX) D U_1 (\bX) & U_1^\dagger (\bX) T (\bX) U_2 (\bX)\\ 
U_2^\dagger (\bX) T^\dagger (\bX) U_1 (\bX) & -U_2^\dagger (\bX) D U_2 (\bX)
\end{array}
\right). 
\end{equation}
Comparing this with (\ref{general superconnection}), 
we find that gauge fields appear
on each of $Dp$-branes and $\overline{Dp}$-branes 
after the gauge transformation, but 
they are pure gauge because 
$-U_1(\bX)^\dagger D U_1(\bX) = -(U_1^\dagger \del_\mu  U_1)(\bX) D\bX^\mu $.
It plays an important role when we discuss 
the gauge flux production in the next section.

The boundary state carries all the information on the system. 
In particular it represents the coupling to any closed string state.
Restricting to the coupling to massless closed string states
and taking appropriate $\alpha' \rightarrow 0$ limit, 
we obtain the low energy effective theory. 
{}From the discussion of the BSFT, the effective action 
for the tachyon field is given by 
\begin{equation}
S[T] =
{}_{\rm NS}\!\bra{0} 
\Tr \hP e^{\int d\hsigma \bM(\hsigma)} 
\ket{Bp}_{\rm NS}, 
\label{mass}
\end{equation}
where the $\bM$ is defined as 
(\ref{tachyon superconnection}). 
It is equivalent to the disk partition function with any number of tachyon 
insertions.
The explicit form of the action is obtained essentially by evaluating the 
bracket of the closed string Hilbert space and performing the 
functional integral in (\ref{Bp}) but it is in general hard task.
For the same number of $Dp$-branes and $\overline{Dp}$-branes ($N=0$),
it is obtained in a closed form in \cite{Kraus:2000nj, Takayanagi:2000rz} 
in the derivative expansion up to ${\cal O}(\del^2T)$.
For $N\ne 0$, there are also some attempts \cite{Jones:2003ae}. 
Similarly, the Chern-Simons term can be written as 
\begin{equation}
S_{\rm CS}[T] =
{}_{\rm RR}\!\bra{C}\Str \hP 
e^{\int d\hsigma \bM(\hsigma)} 
\ket{Bp}_{\rm RR}, 
\label{CS action}
\end{equation}
where $\bra{C}$ is defined as a summation of 
the states of RR-forms, 
\begin{equation}
{}_{\rm RR}\!\bra{C} \equiv \sum_{r} 
{}_{\rm RR}\!\bra{C_{r+1}}. 
\end{equation}
For the RR-sector, the integration over 
the non-zero modes of the bosonic field $X^M$ 
and the fermionic field $\Psi^M$ cancel 
completely because of the worldsheet supersymmetry. 
Therefore the functional integral in  
(\ref{CS action}) reduces to the usual integral 
of the bosonic zero modes $x^\mu$  
and the fermionic zero modes 
play the role of the basis of differential forms $dx^\mu$ 
\cite{Kraus:2000nj,Takayanagi:2000rz}.
Thus the Chern-Simons term (\ref{CS action}) becomes 
\begin{equation}
S_{\rm CS} = T_p \int C \wedge \Str 
e^{2\pi {\cal{F}}}, 
\label{CS-term} 
\end{equation}
where $T_p$ is the tension of a $Dp$-brane 
and ${\cal F}$ is a supercurvature of the superconnection ${\cal A}$ 
on the ($\Z_2$-graded) Chan-Paton bundle on $\R^4$.
For (\ref{tachyon superconnection}), the superconnection ${\cal A}$ and the 
supercurvature ${\cal F}$ are given by
\begin{equation}
{\cal A}= \left(
\begin{array}{cc}
0 & T \\
T^\dagger & 0 
\end{array}
\right), \quad
{\cal F} = \left(
\begin{array}{cc}
-TT^\dagger & \del_\mu T dx^\mu \\
\del_\mu T^\dagger  dx^\mu & -T^\dagger T 
\end{array}
\right).  
\label{A and F}
\end{equation}

\subsection{RR-charges and some simple examples without fluctuations}

We can extract the qualitative feature on the tachyon condensation 
from the boundary interaction or the effective action above. 
First of all, let us consider the conserved RR-charge of $Dp$-branes
after the tachyon condensation.

{}From (\ref{CS-term}), we find the $Dp$-brane charge coupled with
RR $(p+1)$-form is obtained from the 0-form part 
in expansion of $\Str e^{2\pi{\cal F}}$, 
\be
Q_p = \Str e^{-2\pi {\cal A}^2}
={\rm Tr}_{N+M}\, e^{-2\pi TT^\dag} - {\rm Tr}_M\, e^{-2\pi T^\dag T}.
\ee
This just gives ``index'' of the superconnection 
${\cal A}$ in (\ref{A and F}) which is very analogous 
to the index of the Dirac operator.
Indeed, as similar to the index theorem on the Dirac operator, we can easily see
that eigenvalues of $TT^\dag$ and $T^\dag T$ are exactly the same
except for $N$ zero eigenvalues included in $TT^\dag$.
Therefore we find
\be
Q_p = N,
\ee
as expected. Note that the $Dp$-brane charge does not depend on
any detail of the non-vanishing eigenvalues of $T^\dag T$ or $TT^\dag$,
namely, the charge is topologically conserved independent of the fluctuations
of tachyons.


Next, we look at the effective action (\ref{mass}) in the NSNS-sector.
Irrespective of the detailed form of the kinetic term, 
the tachyon potential should take the form,  
\begin{equation}
V(T)= T_p \Tr \exp 2\pi
\left(
\begin{array}{cc}
-TT^\dagger & 0 \\
0 & -T^\dagger T 
\end{array}
\right).
\label{potential}
\end{equation}
For a constant tachyon profile, the effective action (\ref{mass}) 
is exactly given by $S[T]=\int d^{p+1}x V(T)$.
The potential minimum roughly exists at $|T|=\infty$.
For example, if the tachyon has the constant profile as
\begin{equation}
T= \left(
\begin{array}{c}
0 \\
u\,\ {\mathbf 1}_{M}   
\end{array}
\right), \qquad u\in \R,
\label{vacuum solution}
\end{equation}
then the $2M\!\times\! 2M$ part of the potential goes to 
the minimum in the limit $u\to\infty$.
This means that $M$ pairs of $Dp$-branes and 
$\overline{Dp}$-branes decay into the vacuum and $N Dp$-branes remain.
This is also confirmed in terms of the boundary state,  
\begin{equation}
e^{-S_b} \ket{p}_{\rm NS(RR)} 
\rightarrow N \ket{Bp}_{\rm NS(RR)} .
\end{equation}
The final state carries the same value of
the mass and the $Dp$-brane charge proportional to $NT_p$,
thus it is a BPS state.

Now we derive the tachyon condensation to the system of 
$N Dp$-branes and $k D(-1)$-branes, which carry RR $0$-form charge 
proportional to $k T_{-1}$ in addition to the RR $(p+1)$-form charge. 
Since the $D(-1)$-brane is a defect with
the codimension $2n=p+1$ on the $Dp$-branes, 
it is obtained from the $2^{n-1}k$ pairs of $Dp$-branes and 
$\overline{Dp}$-branes by the well-known ABS construction \cite{Witten:1998cd}. 
It is embedded in our tachyon by setting $M=2^{n-1}k$ as
\begin{equation}
T(\bX) = \left(
\begin{array}{c}
0 \\
 -u \bX^\mu
\gamma_\mu \otimes {\mathbf 1}_{k} 
\end{array}
\right),
\label{k D(-1) solution}
\end{equation}
where $\gamma_\mu$ are $2^{n-1}\times 2^{n-1}$
chiral parts of the even dimensional gamma matrices,
\be
\Gamma_\mu=
\left(
\begin{array}{cc}
0 &\gamma_\mu\\
\gamma_\mu^\dag & 0
\end{array}
\right).
\label{higher dim gamma}
\ee
This profile breaks the gauge group down to $U(N)\times U(k)$.
The tachyon potential has the form of a gaussian distribution,
\begin{equation}
V(T)= T_p \Tr \exp 2\pi
\left(
\begin{array}{ccc}
0 & 0 &0 \\
0 & -u^2 |x|^2 \otimes {\mathbf 1}_{2^{n-1}k} &0 \\
0 & 0 & -u^2 |x|^2 \otimes {\mathbf 1}_{2^{n-1}k} \\
\end{array}
\right).
\end{equation}
Suggested by the gaussian nature, 
$2^{n-1}k$ pairs of $Dp$-branes and $\overline{Dp}$-branes disappears
outside the core, $|x|>1/u$, and some defect remains inside the core,
$|x|<1/u$, since the potential does not vanish. 
To show (\ref{k D(-1) solution}) gives the $k D(-1)$ solution
in the effective action, 
now the kinetic term in (\ref{mass}) is also important in addition to the 
potential term, because (\ref{k D(-1) solution}) is space dependent.
Roughly speaking,
both contributions give the correct $\delta$-function as
${\text (potential)}\times{\text (kinetic)} = 
(\frac{\delta^{(4)}(x)}{u^4})\times(u^4)$ 
in the limit $u\to\infty$ \cite{Kraus:2000nj}.
It is indeed seen at the level of boundary states 
by inserting (\ref{k D(-1) solution}) into the matrix $\bM$
(\ref{tachyon superconnection}):
one can show that this system has $N Dp$-brane charge and 
$k$ D(-1)-brane charge in the RR-sector even when $u$ is finite, while
in the NSNS sector bosonic zero-mode part indicates 
that the defect has gaussian distribution 
with the size $|x|\sim 1/u$ and the mass is larger than that of 
$k$ D(-1)-branes \cite{Asakawa:2005vb}.
In the limit $u\to\infty$, we have 
\begin{equation}
e^{-S_b} \ket{Bp}_{\rm NS(RR)} 
\rightarrow N \ket{Bp}_{\rm NS(RR)} + k \ket{B(-1)}_{\rm NS(RR)},
\label{Dp-D(-1)}
\end{equation}
where $\ket{B(-1)}=\ket{\bX^M =0}$ is a boundary state of 
Dirichlet boundary condition for all directions.
Therefore, this gives the tachyon condensation from $2^{n-1}k$ pairs of 
$Dp$/$\overline{Dp}$-branes to $k$ D(-1)-branes at the origin,
while keeping $N Dp$-branes unchanged.

\section{Gauge Flux Production by Unitary Transformations}

The characteristic feature of the system of $(N\!+\!M) Dp$-branes 
and $M \overline{Dp}$-branes is that the tachyon field is 
valued in rectangular matrices. 
The condensation of such a tachyon becomes important in the 
later discussion for the D-brane bound states.
In some cases, the pair of $M Dp$-branes 
and $M \overline{Dp}$-branes decay into the vacuum, 
and the information on the tachyon profile is completely
converted to the gauge field profile on the $N Dp$-branes.
In this section, we describe this mechanism. 

To this end, 
let us consider an arbitrary tachyon profile 
$T(\bX)$ in (\ref{tachyon superconnection}) 
as an $(N+M)\!\times\! M$ complex matrix valued function.  
We assume that it 
is proportional to the parameter $u$.
Roughly speaking, the structure of the tachyon condensation 
is determined by 
the behavior of the tachyon potential (\ref{potential}) 
under the limit $u\rightarrow \infty$. 
Since the tachyon potential is invariant under the gauge symmetry 
$U(N+M)\!\times\! U(M)$, we can choose 
a suitable basis for the trace and rearrange the Chan-Paton indices
so as to separate them into two classes;
$Dp$-branes which may annihilate
and $Dp$-branes which always remain
after the tachyon condensation. 

The combination 
$TT^\dagger (x)$ and $T^\dagger T (x)$ 
appears in (\ref{potential}) are scalar function on $\R^{p+1}$, 
which are valued in hermitian matrices with the sizes of 
$(N+M)\!\times\! (N+M)$ and $M\times M$ and transformed in the 
adjoint representation of $U(N+M)$ and $U(M)$, respectively. 
An immediate consequence is that,
at any point in $\R^4$, $TT^\dagger$ 
has at least $N$ zero eigenvalues and the other $M$ 
eigenvalues completely coincide to those of $T^\dagger T$. 
Therefore, 
we can bring $TT^\dagger$ to the block diagonal form
using a unitary matrix $U_1(x)\in U(N+M)$;
\begin{equation}
U_1^\dagger TT^\dagger U_1 =
\left(
\begin{matrix}
0 & 0 \\
0 & T^\dagger T 
\end{matrix}
\right). 
\label{diagonarize}
\end{equation}
Note that the hermitian matrix $T^\dagger T$ is positive 
semi-definite, that is, 
its all eigenvalues $\lambda_l\, (l=1,\cdots M)$ are non-negative. 
Then, its square-root is also an $M\times M$ hermitian matrix, 
but here the eigenvalues are $\pm \sqrt{\lambda_l}$, that is, 
there is the $2^M$ degeneracy.
By choosing the positive root for all eigenvalues, we 
can define the square-root denoted as  
\begin{equation}
 \lambda \equiv \left(T^\dagger T\right)^{1/2}.
\label{lambda} 
\end{equation} 
The other sign choices are recovered by applying $(\Z_2)^M$ 
transformation, which is a discrete subgroup of $U(M)$.

Furthermore, in this section, 
we assume that the matrix $T^\dagger T$ is strictly positive 
at any point in $\R^4$. 
This guarantees that $\lambda$ has its inverse $\lambda ^{-1}$.  
Under this assumption,
we can easily find the explicit form of $U_1$ as follows.
First, we decompose the matrix $U_1$ by collections of 
$N$ and $M$ vectors in ${\mathbb C}^{N+M}$, respectively, as
\begin{equation}
 U_1 = \left(
V, W
       \right), 
\label{U_1}       
\end{equation}
where $V$ is a collection of $N$ normalized
zero eigenvectors of $T^\dagger$,  
\begin{align}
\label{zero mode condition} 
 T^\dagger V &= 0,  \\ 
 V^\dagger V &= {\mathbf 1}_{M},
 \label{unitary condition 1}
\end{align}
and $W$ is a collection of $M$ normalized
vectors defined by using the existence of $\lambda ^{-1}$ as
\begin{equation}
W=T{\lambda}^{-1}, 
\label{V'}
\end{equation}
since $W$ are arranged so as $TT^\dagger W=W T^\dagger T$. 
Then the conditions for $U_1$ to be unitary matrix, 
\begin{eqnarray}
&&U_1^\dagger U_1
=\left(
\begin{matrix}
 V^\dagger V& V^\dagger T{\lambda}^{-1} \\
 {\lambda}^{-1} T^\dagger V & {\lambda}^{-1} T^\dagger T{\lambda}^{-1}\\
\end{matrix}
	 \right)
=\left(
\begin{matrix}
{\mathbf 1}_{N} & 0 \\
 0 & {\mathbf 1}_{M} \\
\end{matrix}
\right),\\
&&U_1 U_1^\dagger =
V V^\dagger + T{\lambda}^{-2} T^\dagger ={\mathbf 1}_{N+M}, 
\label{unitary condition 2}
\end{eqnarray}
are automatically satisfied because these vectors span the orthonormal basis.
It is also easy to see (\ref{diagonarize}) is satisfied.
Having found a specific unitary matrix $U_1$ (\ref{U_1}), 
there is still a residual symmetry 
$U(N)\!\times\! (\Z_2)^M \subset U(N+M)$, 
which keeps (\ref{diagonarize}) invariant. 
It acts on $U_1$ as 
\begin{equation}
 U_1 = \left(
V, W
       \right) \,\, \rightarrow 
\,\, (VR,WS) \qquad
R\in U(N),\,\, S \in  (\Z_2)^M ,
\label{residual}       
\end{equation}
where the second factor $S$ is the degeneracy stated above.

Some remarks are in order.
In general, both $V$ and $W$ depend on the points in $\R^{p+1}$ 
since $TT^\dagger (x)$ is a function on $\R^{p+1}$.
It is worth to note at this stage 
that the conditions (\ref{zero mode condition}),  
(\ref{unitary condition 1}) and (\ref{unitary condition 2}) 
are the same as the equations appearing in the ADHM construction
(when $p=3$ and $M=2k$). 
The first two conditions 
(\ref{zero mode condition}) and (\ref{unitary condition 1}) means
that $V$ is the normalized zero-modes of the 
``Dirac operator'' $T^\dagger$.
The last one (\ref{unitary condition 2}), called the completeness
relation, says that $V V^\dagger$ and 
$T{\lambda}^{-2} T^\dagger$ are the 
orthogonal projection operator onto the $N$ and $M$ dimensional subspace
of ${\mathbb C}^{N\!+\!M}$, 
respectively. 
We will argue the precise correspondence in the next section.
Note also that $W$ is independent of the parameter $u$ 
because $T\propto u$ and $\lambda ^{-1} \propto u^{-1}$. 
Therefore, the gauge transformation $U_1$ is also independent of $u$.

What we have done is shuffling the $(N\!+\!M) Dp$-branes and relabelling 
the Chan-Paton indices such that
the tachyon and its potential has the form, 
\begin{align}
T'=U_1^\dagger T=\left(
\begin{matrix}
0\\
\lambda \\
\end{matrix}
\right), \qquad
V(T')=T_3 \, \Tr
\exp\left(
\begin{matrix}
0 & 0 & 0 \\
0 & -\lambda^2 & 0  \\
0 & 0 & -\lambda^2 \\
\end{matrix}
	 \right). 
\label{transformed tachyon}
\end{align}
Now this tachyon profile is to be compared with that of
 (\ref{vacuum solution}) and (\ref{k D(-1) solution}). 
Although $\lambda ^2(x)$ depends on the space coordinate, their eigenvalues 
are strictly positive at any point in $\R^{p+1}$ so that
the tachyon potential behaves similarly as (\ref{vacuum solution}).
Then, under this condition,
it is expected that the $M$-pairs of $Dp$-branes and
$\overline{Dp}$-branes completely disappear and
$N Dp$-branes keep alive in the limit of $u\to\infty$.
If we want to know more details of the tachyon condensation,
we should also evaluate the kinetic term not only the potential term
as for the case of (\ref{k D(-1) solution}) 
since the tachyon profile depends on the spacetime coordinate. 
Unfortunately, however, it is quite hard task in general since 
we do not know the explicit form of the effective action.

So far we restrict the discussion only on the tachyon potential
in order to sketch the behavior of the system.
However, we can treat the system more precisely using the boundary state
formalism. 
The main difference is that 
the gauge invariant object is now the boundary interaction 
$e^{-S_b}$ (\ref{boundary interaction})
and the unitary transformation induces 
inevitably the pure gauge connection (\ref{gauge transformation}) 
in the boundary interaction.
We now look at the same gauge transformation at the level of boundary state 
and see how the (non-trivial) gauge field appears. 
Let us consider the boundary interaction with 
$\bM$ (\ref{tachyon superconnection}).
Among the gauge symmetry $U(N+M)\!\times\! U(M)$, 
the gauge function (\ref{gauge function}) of our concern is given by  
\begin{equation}
 U(\bX)\equiv \left(
\begin{matrix}
U_1 (\bX) & 0 \\
 0 & {\mathbf 1}_{M} \\
\end{matrix}
	 \right),
 \label{U(X)}
\end{equation}
where $U_1 \in U(N\!+\!M)$ is given by (\ref{U_1}). 
Note that it is a functional of
the string (super)coordinate $\bX^\mu(\hsigma)$.
Then the corresponding gauge transformation (\ref{gauge transformation})
for $\bM$ is 
\begin{equation}
 \bM \mapsto \bM' = \bP + \bA, 
 \label{mapped superconn}
\end{equation}
where
\begin{equation}
 \bP \equiv  U^\dagger \bM U = \left(
\begin{matrix}
 0 & 0 & 0 \\
 0 & 0 & \lambda  \\
 0 & \lambda  & 0 
\end{matrix}
 \right), \qquad
 \bA \equiv -U^\dagger DU = \left(
\begin{matrix}
 -V^\dagger D V & -V^\dagger D W & 0 \\
 -W^{\dagger} D V & -W^{\dagger}DW
 & 0 \\
 0 & 0 & 0 
\end{matrix}
	    \right).
 \label{mapped P and A}
\end{equation}
The first term $\bP$ comes from the adjoint action of $U(\bX)$ acting on $\bM$.
This term reduces to the transformation law 
(\ref{transformed tachyon}) in the effective theory.
The second term $\bA$ is the induced pure gauge connection by 
the transformation $U(\bX)$. 

Recalling (\ref{general superconnection}),
$\bA$ represents an insertion of the massless gauge fields 
on the $(N\!+\!M) D3$-branes.
Initially, the gauge transformation is introduced to block-diagonalize 
the tachyon potential, 
but it also decomposes $\bM'$ into a tachyon part $\bP$ and a gauge field
part $\bA$. 
It says that not all the component of the initial tachyon profile
are indeed tachyonic.
Note also that $\bP$ is $u$-dependent while $\bA$ is $u$-independent.
In general, only the $u$-dependent part is 
responsible for the tachyon condensation 
$u\rightarrow\infty$, while the others are regarded as fluctuations 
around the condensed vacuum. 
Such a decomposition is not only for the convenience 
but also relies on the physical interpretation of the system
after the tachyon condensation.
There is a case where we can further
extract $u$-independent parts from $\bP$ which are regarded as
fluctuations, 
but it heavily relies on the detailed form of the tachyon profile.
We will come back to this point in  the section 4.5.
Hence in this section, we simply treat $\bP$ as the principal part of the 
tachyon condensation (background) and $\bA$ as a fluctuation.
Correspondingly, in evaluating the boundary interaction, 
it is convenient to use the following identity for the 
(super)path-ordered product
(for the proof see the appendix \ref{app formula}), 
\begin{align}
 \hP e^{\int d\hsigma \bM(\hsigma)}
 &= \hP e^{\int d\hsigma (
 \bP(\hsigma))+\bA(\hsigma))} \nn \\
 &\equiv
 \hP e^{\int d\hsigma \bP(\hsigma)}
 +\sum_{n=1}^\infty \int d\hsigma_1 \cdots d\hsigma_n
 \bA(\hsigma_n)
 {\mathbf T}_{\bP}(\hsigma_n-\hsigma_{n-1})
 \bA(\hsigma_{n-1}) \nn \\
 & \hspace{5cm}
 \times{\mathbf T}_{\bP}(\hsigma_{n-1}-\hsigma_{n-2})
 \cdots {\mathbf T}_{\bP}(\hsigma_{2}-\hsigma_1)
 \bA(\hsigma_1),
 \label{separation formula}
\end{align}
where the ``transfer matrix'' ${\mathbf T}_{\bP}(\hsigma'-\hsigma)$ is 
defined only through $\bP$ as 
\begin{equation}
 {\mathbf T}_{\bP}(\hsigma'-\hsigma)\equiv
  \hP e^{\int d\hsigma''\bP(\hsigma'')
  \left(
  \Theta(\hsigma'-\hsigma'') - \Theta(\hsigma''-\hsigma)
  \right)}.
  \label{hP-transfer matrix}
\end{equation}
This identity means that inserting infinitely many numbers 
of the vertex operators $\bP$ first, which fixes the background,
and then inserting $\bA$ as perturbations.
${\mathbf T}_{\bP}(\hsigma'-\hsigma)$
is further
estimated by expanding $\bP$ with respect to $\theta$ as 
$\bP (\hsigma)= P_0 (\sigma)+ \theta P_1 (\sigma)$ 
and again using the similar identity for the standard path-ordering, 
\begin{align}
{\mathbf T}_{\bP}(\hsigma'-\hsigma)
 &= P e^{\int_\sigma^{\sigma'} d\sigma''
 (P_1(\sigma'')-P_0^2(\sigma''))} \nn \\
 &=
 T_{-P_0^2}(\sigma'-\sigma)
 \nn \\
 &\hspace{1cm}
 +\sum_{n=1}^\infty \int d\sigma_1 \cdots d\sigma_n
 T_{-P_0^2}(\sigma'-\sigma_{n})
 P_1(\sigma_n)
 T_{-P_0^2}(\sigma_n-\sigma_{n-1}) 
 P_1(\sigma_{n-1})  \nn \\
&\hspace{3.7cm}
 \times T_{-P_0^2}(\sigma_{n-1}-\sigma_{n-2})
 \cdots
 {T}_{-P_0^2}(\sigma_{2}-\sigma_1)
 P_1(\sigma_1)
 {T}_{-P_0^2}(\sigma_{1}-\sigma), 
 \label{separation formula 2}
\end{align}
where 
\begin{align}
 T_{-P_0^2}(\sigma'-\sigma) &\equiv 
 Pe^{-\int_\sigma^{\sigma'}d\sigma'' P_0^2(\sigma'')}
 \nn \\
 &=\left(
\begin{matrix}
 1 & 0 & 0 \\
 0 & Pe^{-\int d \sigma \lambda^2 (X)} & 0 \\
 0 & 0 & Pe^{-\int d \sigma \lambda^2 (X)}
 \end{matrix}
		     \right).  
\label{P-transfer matrix}
\end{align}
The main point is that 
$\lambda^2 $ diverges uniformly (irrespective of $X(\sigma)\,$)  
in the limit of $u\to\infty$
since we have assumed that the matrix $\lambda$ is strictly positive definite
and also that all the eigenvalues are $u$-dependent.
Hence, it is concluded that (\ref{P-transfer matrix}) reduces
to the projection operator onto the $N$-dimensional subspace
of the $(N+M)$-dimensional Chan-Paton space in the limit $u\to\infty$:
\begin{equation}
  T_{-P_0^2}(\sigma'-\sigma)
  \stackrel{u\to\infty}{\longrightarrow} \left(
\begin{matrix}
{\mathbf 1}_{N} & 0 \\
 0 & 0 
\end{matrix}
					\right)
  \equiv \cP_N. 
\end{equation}
Then, it is easy to see that
(\ref{hP-transfer matrix}) also goes to the same projection
operator $\cP_N$ under the limit $u\rightarrow\infty$. 
This is because each $P_1(\sigma)$ in (\ref{separation formula 2}) 
is now sandwiched by $\cP_N$ but it vanishes ($\cP_N P_1(\sigma)\cP_N =0$).
The fact that the condensation of 
the principal part $\bP$ becomes the projection operator $\cP_N$
supports the rough sketch of the behavior obtained from the tachyon potential.
But we emphasize that this is true only under the assumption that 
whole $\bP$ is relevant for the tachyon condensation. 

We can further estimate 
(\ref{separation formula}) in the limit of $u\to\infty$ as 
the same manner:
each $\bA(\hsigma)$ are replaced by $\cP_N \bA(\hsigma)\cP_N $ and 
we obtain
\begin{align}
 \hP e^{\int d\hsigma \bM(\hsigma)}
 &\stackrel{u\to\infty}{\longrightarrow}
 \cP_N \left(\hP e^{\int d\hsigma \bA(\hsigma)}\right) \cP_N \nn \\
&=\hP \exp\left\{
\int d\hsigma \left(
\begin{matrix}
-V^\dagger D V & 0 \\
 0 & 0
\end{matrix}
 \right)
\right\}. 
\end{align}
Therefore, in the limit $u\rightarrow \infty$, 
the boundary state reduces to
\begin{equation}
e^{-S_b} \ket{Bp}_{\rm NS(RR)} 
\rightarrow {\rm (S)Tr}_{N} \left(\hP e^{-\int d\hsigma V^\dagger D V}\right)
\ket{Bp}_{\rm NS(RR)}.
\label{final product}
\end{equation}
This means that 
the $M$ pairs of the $Dp$-branes and
$\overline{Dp}$-branes completely disappear and only
$N Dp$-branes remain after the tachyon condensation
($u\to\infty$).
Moreover, an $U(N)$ gauge connection, 
\begin{equation}
 A_\mu = -V^\dagger \del_\mu V,
  \label{gauge field}
\end{equation}
appears on the remaining $N Dp$-branes  
since $V^\dagger D V = (V^\dagger \del_\mu V) D\bX^\mu$
(see (\ref{general superconnection})).
The residual gauge symmetry $U(N)$ in (\ref{residual}) remains
on the $N Dp$-branes, which act on (\ref{gauge field}) as
\begin{equation}
 A_\mu \mapsto A'_\mu = R^\dagger \del_\mu R + R^\dagger A_\mu R .
\end{equation}

We have given the mechanism 
that the initial information on the tachyon profile 
is converted to the gauge profile (\ref{gauge field}) 
on the remaining $N D3$-branes.
Since it is simply a gauge transformation, it gives 
a unitary equivalence between the $Dp/ \overline{Dp}$-system with 
a tachyon and the $Dp$-branes with a flux.
We emphasize that this mechanism is the consequence of the 
(non-vanishing) tachyon between the $N Dp$-branes and 
$M Dp/ \overline{Dp}$-branes.
In fact, for the tachyon profiles (\ref{vacuum solution}) 
and (\ref{k D(-1) solution}), we can not apply the
transformation (\ref{diagonarize}) (or it is trivial).
As we will see in the next section, it is the essence of the 
bound state formation, where (\ref{gauge field}) corresponds 
to the $U(N)$ gauge connection in the ADHM construction. 
However, since it is pure gauge form, 
it seems to make no physical effect at first sight.
In fact, at the level of the Chern-Simons term (\ref{CS-term}), 
the supercurvature transforms covariantly under $U(x)$ 
so that the non-trivial curvature seems not to be produced by 
such a gauge transformation.
But it is not true in general.
The point is that the unitary transformation could be 
{\it a large gauge transformation}, and moreover, 
even if the total curvature in the $U(N+M)$ 
is still trivial, the curvature projected onto the $U(N)$ part may be 
non-trivial, while the other part of $U(N+M)$ becomes 
hidden under the tachyon condensation.
Whether (\ref{gauge field}) is in fact non-trivial or does not 
depend on the topological information carried by 
the gauge function $U(\bX)$, or equivalently, 
the initial tachyon profile. 

Before concluding this section, we give a few remarks.
First of all, although we have used the transformation of the form 
(\ref{diagonarize}), the argument in this section is unchanged 
if the $N$ zero-modes $V$ are successively separated. 
In other words, a further transformation of (\ref{transformed tachyon}) 
by the $U(M)\times U(M)$ gives the same conclusion, 
because the induced pure gauge connection by this transformation 
is projected out by ${\cal P}_N$.
Secondary, the argument relies only on the existence of the 
gauge transformation and the appropriate separation of the principal 
part.
So it is also applicable to a case with initial 
non-vanishing massless fields by an appropriate generalization.
In this case, they are not relevant for the 
tachyon condensation itself but we should be careful about the 
gauge symmetry breaking by them.
Finally, if $T^\dagger T$ is not always positive, 
there should be zero modes at least at some space point. 
Even in this case, by separating zero-mode part and considering only the small 
matrix, a rearrangement such as (\ref{diagonarize}) still 
works. As a consequence, we obtain after the transformation the system 
of $(N+M') Dp$/$M' \overline{Dp}$-branes with a flux, where 
$M'(< M)$ is the number of the above zero-modes.

\section{Instantons in the $D3$/$\overline{D3}$ system}

In the previous section,
we considered the system of coincident $(N+M) Dp$-branes
and $M \overline{Dp}$-branes with a generic tachyon profile. 
In this section, 
we concentrate on the case of $(N+2k) D3$-branes and
 $2k \overline{D3}$-branes with tachyon profiles, 
which carry the $k D(-1)$-brane charge.
We will show the physical equivalence between a system of  
$k D(-1)$-branes and $N D3$-branes with open strings stretched
between them and a system of $N D3$-branes 
with a gauge flux through a unitary equivalence in the 
$D3$/$\overline{D3}$ system.
This in particular explains the ADHM construction 
of the instantons in $4D$ gauge theory.

\subsection{Two pictures of bound states}

Let us first recall the fact about the $D(-1)$-$D3$ bound states 
described in the section \ref{Intro} in slightly more detail. 
It is well known that 
the system of well separated $N D3$-branes and $k D(-1)$-branes
preserves the spacetime supersymmetry.
Furthermore, we can consider the collective excitations of them by 
turning on massless open strings.
If $N D3$-branes and $k D(-1)$-branes are located closely with 
each other, stretched open strings between them become massless and
the system forms a BPS bound state.
It is known that this system is marginal bound state, 
that is, the total mass of the bound state 
equals to that of the individual 
$N D3$-branes and $k D(-1)$-branes. 
In the section \ref{Intro}, we present two physically equivalent ways
to describe this BPS bound state
in the superstring theory  
(see Fig.~\ref{strategy} in the section \ref{Intro}).

The first picture (a) is the system of coincident $N D3$-branes
and $k D(-1)$-branes and massless open strings, whose ends
are attached at least on the $D(-1)$-branes (DN- and DD-strings). 
Such open strings are the collective degrees of freedom that survive 
in the low energy limit of $\alpha'\to 0$ with fixing
the $0$-dimensional gauge coupling $g_0^2\sim g_s \alpha'^{-2}$.
Here the excitations on $D3$-branes become extremely heavy 
and other massive excitations or gravitational interactions are 
decoupled.
Then 
the effective theory on the $D(-1)$-branes
becomes the $0$-dimensional supersymmetric gauge
theory with gauge group $U(k)$ and flavor symmetry $U(N)$. 
The (bosonic) matter contents of 
this effective theory are in the following. 
{}From the massless excitation of open strings stretched
between $D(-1)$-branes, 
there appear $10$ scalar fields $\Phi^M\, (M=0,1,\cdots,9)$
in the adjoint representation of $U(k)$, 
which describe the transverse positions of $D(-1)$-branes.
Open strings between $D(-1)$-branes and $D3$-branes 
give scalar fields $H=(I^\dagger,J)$ where $I$ and $J$ are 
in the $(\bar{N},{k})$ and $(N,\bar{k})$ representation
of $U(N)\times U(k)$, respectively.

In terms of low energy effective theory,
the BPS condition of the system is equivalent to the 
condition where the vacuum expectation value (vev) of 
matter fields lie in the supersymmetric vacua.
The vacua of the theory have several branches. 
The Higgs branch,
where the vev of the scalar fields $\Phi^i \,(i=4,\cdots,9)$ are zero 
while $H$ and $\Phi^\mu , (\mu=0,1,2,3)$ have non-zero vev, 
corresponds to the situation where the $D(-1)$-branes are 
located inside the $D3$-branes.
The Coulomb branch,
where $H=0$ and $\Phi^M\ne 0$,  
describes the situation in which the $D(-1)$-branes move off the 
$D3$-branes.
There are also mixed branches where some of the $D(-1)$-branes 
still lie on the $D3$-branes and other $D(-1)$-branes
move off the $D3$-branes.
If the field configurations lie on these flat directions,
the system has the same mass as that of $N D3$-branes
and $k D(-1)$-branes.

Another picture (b) is the system of 
$N D3$-branes with massless open strings excited on them
(NN-strings).
In this picture, the $k D(-1)$-branes are completely 
dissolved into the $N D3$-branes as a gauge field configuration.
Therefore, this description is quite well compatible with instantons  
in the supersymmetric gauge theory. 
In fact, in the low energy limit $\alpha' \to 0$
with fixing the $4$-dimensional gauge coupling 
$g_{4}^2 \sim g_s$, 
the effective theory on the $N D3$-branes 
becomes the ${\cal N}=4$ $U(N)$
supersymmetric Yang-Mills theory with the Chern-Simons term,
where the $k D(-1)$-branes charge is carried by 
the $k$-instanton gauge configuration through the coupling
$C\int \Tr(F\wedge F)$.
If we restrict ourselves to the vev for the $6$ transverse scalar 
fields $\Phi^i$ to be zero, the bosonic content of this theory
is simply given by the $U(N)$ Yang-Mills theory with a $\theta$-term.
Then the BPS condition (vacua) 
is equivalent to the self-dual field strength,
where the Yang-Mills action is equal to the $k$-instanton action.

In the Higgs branch of the picture (a) and in the $k$-instanton sector 
of the picture (b), 
the mass (action) and the RR-charges of the BPS configurations 
are the same so that they are thought to be equivalent.
However, this equivalence seems to be mysterious 
since they are defined in the very different low energy limit of the 
seemingly different D-brane systems with each other.
Nevertheless,
a remarkable correspondence between two pictures at low energy 
is provided by the ADHM construction 
\cite{Atiyah:1978ri, Corrigan:1983sv}.
In this construction, 
the D and F-flatness condition defining the Higgs branch  
(ADHM equation) in the picture (a) is mapped to the self-dual condition 
in the picture (b).
The instanton moduli space ${\cal M}_{N,k}$ in the picture (b) 
is also recovered by the ADHM data, 
$H$ and $\Phi^\mu$, divided by the auxiliary symmetry $U(k)$,
\begin{equation}
 {\cal M}_{N,k}
 = \{(H,\Phi^\mu)|\mbox{ADHM eqs.}\}/U(k). 
 \label{moduli space}
\end{equation}
Moreover, singularities in the instanton moduli space ${\cal M}_{N,k}$,
where we cannot construct a regular instanton gauge field 
(small instanton singularity), 
correspond to the conical singularities in 
the hyper-K\"ahler quotient (\ref{moduli space}).
In the D-brane setting, this limit connects different branches in (a) 
and it is regarded as a stringy resolution of the small instanton singularity.
This is the ordinary explanation of the ADHM construction 
in terms of the dynamics of D-branes in string theory.
 
In this paper,
we show that the correspondence between the above two BPS systems
is understood as a gauge equivalence. 
We also propose that the equivalence between (a) and (b) 
is not restricted on the low energy limits nor on the BPS configurations.
Our strategy is summarized in the figure \ref{strategy} in the section \ref{Intro}.
First, we consider the D-brane system in the both pictures, 
where the only excitations are the low energy degrees of freedom.
They are described in the full string theory 
by the boundary states with specific open string excitations, 
that is, $k D(-1)$-branes and $N D3$-branes with 
DN- and DD-strings in (a) 
and $N D3$-branes with NN-strings in (b). 
Next, each boundary state is obtained from the same system of 
$(N+2k) D3$-branes and $2k \overline{D3}$-branes as a result 
of the tachyon condensation.
In the picture (a) $2k$ pairs of $D3$/$\overline{D3}$-branes give 
$k D(-1)$-branes, while in the picture (b) they are annihilated 
into the closed string vacuum.
In the following subsections, we will construct the bound state of $N$
$D3$-branes and $k D(-1)$-branes in the both pictures using the
tachyon condensation from a $D3$/$\overline{D3}$ system
and we will show that the two descriptions are 
simply related by the unitary gauge transformation in 
$U(N+2k)\times U(2k)$, 
which is the mechanism described in the section 3.
Therefore, two pictures arise as specific gauges and 
they are unitary equivalent with each other as the D-brane systems.

\subsection{ADHM data as tachyon profile}

In this subsection, 
we describe the picture (a), that is, 
a $k D(-1)$/$N D3$ bound state with 
open string excitations $H$ and $\Phi^\mu$ 
in terms of boundary state. 
The boundary states of individual $D3$-branes 
and $D(-1)$-branes is a linear combination of them 
as given by (\ref{Dp-D(-1)}) with $p\!=\!3$. 
Excitations of $4$ scalar fields $\Phi^\mu$ are simply incorporated by 
the boundary interaction $e^{-S_b}$ acting on $\ket{B(-1)}$, which 
represents arbitrary number of boundary insertions of 
massless vertex operators on the 
disk with the Dirichlet boundary condition. 
More precisely, it has the form, 
\begin{equation}
 e^{-S_b}={\rm (S)Tr}_k \, \hP e^{-i\int d\hsigma \Phi^\mu
  \bP_\mu(\hsigma)}. 
 \label{scalar}
\end{equation}
On the other hand, since $H$ is DN-strings
connecting the $D3$-branes and $D(-1)$-branes,
their insertions should be accompanied with
the boundary condition changing operators 
or twist fields in general,
which is not well developed
so far to handle in the context of boundary state.%
\footnote{
For calculations of the 
mixed disks along this line, see \cite{Billo:2002hm} and references their in.
}
However, there is another way to represent them:
we come back to the system of $(N+2k) D3$-branes
$2k \overline{D3}$-branes and add the fluctuations
corresponding to $H$ and $\Phi^\mu$ around the $k D(-1)$-brane
solution (\ref{k D(-1) solution}).
Explicitly, it corresponds to the tachyon profile, 
\begin{equation}
 T(\bX)=u\left(
\begin{array}{c}
 H \\
 (\Phi^\mu - \bX^\mu)\otimes \sigma_\mu
\end{array} 
    \right)
 =u\left(
\begin{matrix}
 I^\dagger & J  \\
 {B}_2^\dagger - {{\mathbf Z}}_2^\dagger &
 -{B}_1 + {{\mathbf Z}}_1 \\
 {B}_1^\dagger - {{\mathbf Z}}_1^\dagger &
 {B}_2 - {{\mathbf Z}}_2
\end{matrix}
 \right),
\label{ADHM data}
\end{equation}
where $T$ is valued in  $(N\!+\!2k)\times 2k$ matrix, 
$H$ is a complex $N\times 2k$ matrix, 
$\Phi_\mu \, (\mu=0,1,2,3)$ are hermitian
$k\times k$ matrices, respectively.
$\sigma_\mu$ are quotanion basis defined by
$\sigma_\mu=(1,-i\tau_i)$ using Pauli 
matrices $\tau_i\, (i=1,2,3)$.
In the second expression of (\ref{ADHM data}),
we have introduced a complex notation 
where ${B}_{1,2}$ and ${{\mathbf Z}}_{1,2}$ are defined as
\begin{equation}
 {B}_1 \equiv \Phi^2+i\Phi^1, \qquad 
  {B}_2 \equiv \Phi^0+i\Phi^3,  
\end{equation}
and
\begin{equation}
 {{\mathbf Z}}_1 \equiv \bX^2+i\bX^1, \qquad 
 {{\mathbf Z}}_2 \equiv \bX^0+i\bX^3,
\end{equation}
respectively.
In this setting, $\Phi^\mu$ describe
the degrees of freedom of open strings
between $2k D3$-branes and $2k \overline{D3}$-branes. 
When $H=0$, the tachyon condensation of this profile 
(\ref{ADHM data}) gives the boundary interaction
(\ref{scalar}) as expected \cite{Asakawa:2002ui} (see also the section 4.5). 
This means that $\Phi^\mu$ become the transverse scalar fields on 
$k D(-1)$-branes after the tachyon condensation. 
Similarly, $H$ in (\ref{ADHM data}) corresponds to the open strings between 
the $N D3$-branes and the $2k$ pairs of $D3$/$\overline{D3}$-branes. 
Then it is natural to expect that it would describe
the DN-strings between $D3$-branes
and $D(-1)$-branes after the tachyon condensation. 

This tachyon profile (\ref{ADHM data}) is evidently 
proportional to the ``Dirac operator'' $\Delta$ in the canonical form 
in the context of the ADHM construction
(see e.g. \cite{Corrigan:1983sv}), 
that is, $H$ and $\Phi^\mu$ are identified with
the ADHM data and $T=u\Delta$.
Such an identification is first appeared in the literature 
\cite{Akhmedov:2001jq} to describe $D5$-branes inside the $D9$-branes
in the context of the effective action.  
They have evaluated the Chern-Simons term 
(\ref{CS-term}) in the simplest example of $H$ and $\Phi^\mu$ ($N=2,\,k=1$)
and confirmed that (\ref{ADHM data}) possesses the correct RR-charge density
expected from the instanton configuration.
However, the ADHM equation (or D and F-flatness condition) itself cannot 
recovered only by the Chern-Simons term. 
In terms of boundary states, 
the BPS condition are seen by ``mass = RR-charge''
by evaluating both the BSFT action (\ref{mass}) 
and the Chern-Simons term (\ref{CS action}).
In this way the ADHM equation would be recovered in the 
boundary state formalism essentially.
However, as emphasized in \cite{Hashimoto:2005qh}, it is difficult task
to see whether the ADHM equation is the exact BPS condition which
receives no $\alpha'$-corrections.%
\footnote{
The authors would like to thank to K.~Hashimoto and S.~Terashima for
discussing this point. 
}
Thus we will not analyze along this line in this paper,
but later we will encounter how the ADHM equation is also in a sense
special in the full special in the full string theory as in the ordinary
ADHM construction. 

In summary, we demonstrated a realization of the 
$N D3$-$k D(-1)$ bound state
in terms the boundary state of $D3$/$\overline{D3}$ system, 
which corresponds to the picture (a).
Here the RR $0$-form charge is carried by the $k D(-1)$-branes and
the information on the instanton moduli is contained
in the massless excitations of open strings on the $D(-1)$-branes. 
These excitations are already known as $N D3$-$k D(-1)$ bound state 
realization of the ADHM data but 
we stress that both the D-brane configuration
and the open string fluctuation on them 
are packed together in the tachyon profile (\ref{ADHM data}) 
exactly in the ADHM matrix form.

\subsection{ADHM construction as a gauge transformation}
\label{ADHM as gauge}

We next consider another picture (b) of the same system 
which can be directly compared with the instantons
of the $4D$ gauge theory in the low energy limit. 
We can easily write down the boundary state corresponding to the picture
(b) as
\begin{equation}
 {\rm (S)Tr}_N \hP e^{\int d\hsigma A_\mu(\bX) D \bX^\mu}
  \ket{B3}_{\rm NS(RR)},
  \label{boundary state in the picture b}
\end{equation}
where $A_\mu$ is the gauge flux.
However, we would like to construct it using the tachyon condensation in
order to compare it with the boundary state in the picture (a)
constructed above. 
To achieve it,
$2k$ pairs of $D3$-branes and $\overline{D3}$-branes
should be disappeared into the vacuum completely and
a field configuration on the $N D3$-branes should possesses
all the information on the $k D(-1)$-branes. 
{}From now on, we show that, at the level of the full string theory,
this picture (b) is obtained 
from the picture (a) via the mechanism discussed in the section 3,
that is, via a gauge transformation.
We will see that under the same assumption of the ADHM construction, 
this process is nothing but the stringy realization of that construction.

We start with the tachyon profile (\ref{ADHM data}).
One of the working assumption in section 3 is 
the positivity of the $2k\times 2k$ matrix $T^\dagger T$.
So we examine the properties of $T^\dagger T$ for the tachyon profile
(\ref{ADHM data}), which is written 
in terms of $H$ and $\Phi^\mu$ as 
\begin{align}
T^\dagger T
&= u^2 \left\{H^\dagger H +\overline{\sigma}_\mu \sigma_\nu 
(\Phi^{\mu}-\bX^\mu )(\Phi^\nu-\bX^\nu ) \right\} \nn \\ 
&= u^2 \left\{H^\dagger H +
(\Phi^{\mu}-\bX^\mu )(\Phi_\mu-\bX_\mu ) 
+i\eta_{\mu\nu}^{(-)i}\tau_i [\Phi^{\mu},\Phi^{\nu}]
\right\}.
 \label{lambda^2-pre}
\end{align}
In the second line, we used the hermiticity of 
$\Phi^\mu$ and the following identities for the quotanion basis, 
\begin{align}
\begin{array}{l}
 \sigma_\mu \bar{\sigma}_\nu = 
 \delta_{\mu\nu} {\mathbf 1}_{2}
 + i\eta_{\mu\nu}^{(+)i}\, \tau_i , \\
 \bar{\sigma}_\mu {\sigma}_\nu
 =
 \delta_{\mu\nu} {\mathbf 1}_{2}
 + i\eta_{\mu\nu}^{(-)i}\,\tau_i\,,
\end{array}
 \label{identity-1}
\end{align}
where $\eta_{\mu\nu}^{(\pm)i}$ are so-called 't Hooft's symbols defined by
\begin{equation}
 \eta_{\mu\nu}^{(\pm)i} \equiv
  \epsilon_{0 i\mu\nu} \mp \delta_{i\mu}\delta_{\nu 0}
  \pm \delta_{i\nu}\delta_{\mu 0},  
  \label{tHooft symbol}
\end{equation}
which satisfy the (anti-)self-dual relations, 
\begin{equation}
 \epsilon_{\mu\nu\rho\sigma}\eta_{\mu\nu}^{(\pm)i}
  = \pm \eta_{\mu\nu}^{(\pm)i}. 
  \label{self-duality of tHooft symbol}
\end{equation}
By expanding also $H^\dagger H$ with respect to Pauli matrices, 
$T^\dagger T$ can be decomposed as 
\begin{align}
 T^\dagger T &=
 u^2  {\mu_0}\otimes {\mathbf 1}_{2}
 + u^2 \sum_{i=1}^3 \mu_i \otimes \tau_i,
  \label{expansion of TT}
\end{align}
where $\mu_0$ and $\mu_i\,(i=1,2,3)$ are $k\times k$ hermitian matrices, 
which are written in the complex notation as
\begin{align}
 2{\mu_0} &\equiv II^\dagger + J^\dagger J
 + \frac{1}{2}\sum_{a=1}^2 \left\{
 B_a - {\mathbf Z}_a, B_a^\dagger - {\mathbf Z}_a^\dagger
 \right\}, 
 \label{energy}\\
 \mu_{\R} &\equiv 2 \mu_3 
  = II^\dagger - J^\dagger J
  +\left[{B}_1, {B}_1^\dagger\right]
  +\left[{B}_2, {B}_2^\dagger\right], 
 \label{real momentum}\\
 \mu_{\mathbb C} &\equiv \mu_1 + i \mu_2 
  = IJ + \left[{B}_1, {B}_2 \right].
 \label{complex momentum}
\end{align}
Notice that only ${\mu_0}$ depends on $\bX^\mu$
in these expressions. 

We have identified tachyon profile with the ADHM matrix as $T=u\Delta $.
The assumption of the ADHM construction is to impose
the ADHM equation (condition),
\begin{equation}
 \mu_{\R} = \mu_{\mathbb C} = 0,
  \label{ADHM constraint}
\end{equation}
and the invertiblity of $\Delta ^\dagger \Delta$.
They are equivalent to the (strict) positivity of 
$T^\dagger T=u^2{\mu_0}\otimes {\mathbf 1}_{2}$ in (\ref{expansion of TT}).
Then we can safely apply the mechanism 
demonstrated in the section 3 since the working assumption is
automatically satisfied.

Under these assumption, we can show that the ADHM construction is
naturally reproduced. 
First, note that $\lambda = u(\mu_0)^{1/2}\otimes {\mathbf 1}_{2}$ 
(\ref{lambda}) is 
invertible so that 
the unitary matrix $U_1\in U(N+2k)$ in (\ref{U_1}) is well-defined. 
In particular, (\ref{zero mode condition}) and
(\ref{unitary condition 1}) 
are the normalized zero-mode condition for the zero-dimensional 
``Dirac operator'', 
\begin{align}
\label{inst zero mode condition}
 \Delta  ^\dagger V &= 0,  \\ 
 V^\dagger V &= {\mathbf 1}_{N},
 \label{inst unitary condition 1}
\end{align}
and (\ref{unitary condition 2}) reduces to the completeness relation, 
\begin{align}
V V^\dagger + \Delta ({\mu_0}^{-1}\otimes {\mathbf 1}_{2}) 
\Delta^\dagger ={\mathbf 1}_{N+2k}, 
 \label{inst unitary condition 2}
\end{align}
of the ADHM construction, respectively.
Under the gauge transformation $U\in U(N+2k)\times U(2k)$ in (\ref{U(X)}), 
the tachyon and its potential are transformed as
\begin{align}
T'=\left(
\begin{matrix}
0\\
u(\mu_0)^{1/2}\otimes {\mathbf 1}_{2} \\
\end{matrix}
\right), \qquad
V(T')=T_3 \, \Tr
\exp\left(
\begin{matrix}
0 & 0 & 0 \\
0 & -u^2 \mu_0 \otimes {\mathbf 1}_{2} & 0  \\
0 & 0 & -u^2 \mu_0 \otimes {\mathbf 1}_{2} \\
\end{matrix}
	 \right). 
 \label{ADHM tachyon}
\end{align}
As noted in the section 3, we can further diagonalize each $\mu_0$ 
by the unitary transformation $U(k)$, which is the diagonal subgroup of
$U(k)\times U(k)\subset U(N+2k)\times U(2k)$. 
Then we can see that this profile sits at the minimum of the potential 
in the limit $u\rightarrow \infty$, 
because each eigenvalue of $\mu_0$ is strictly 
positive and this behavior is independent of $\bX$.
Therefore, $2k$ pairs of $D3/\overline{D3}$-branes are expected 
to annihilate 
into the vacuum.
We emphasize that this feature is due to the nonzero $I$ and $J$, 
that is, originally the excitation of the stretched DN-string between 
$D(-1)$-branes and $D3$-branes.
At the level of boundary states, 
the boundary interaction $e^{-S_b}$ is given by 
(\ref{mapped superconn}) and (\ref{mapped P and A}). 
In particular, the pure gauge connection $\bA$ is induced in the 
$U(N+2k)$ part.
On the other hand, 
the relevant part in $\bP$ is given by 
$T'$ in (\ref{ADHM tachyon}) 
with diagonalized by $U(k)$ as above.  
Although the additional pure gauge connection in the $U(k)$ 
part is also introduced by this transformation, 
it becomes hidden under the condensation $u\rightarrow \infty$, 
which is exactly the same as the evaluation in section 3.
This explains the residual $U(k)$ symmetry in the ADHM construction, 
which cannot be seen after the unitary rotation.
As a result, $2k$ pairs of $D3/\overline{D3}$-branes disappear and 
we obtain the boundary state for $N D3$-branes as
\begin{equation}
e^{-S_b} \ket{B3}_{\rm NS(RR)} 
\rightarrow {\rm (S)Tr}_{N} \left(\hP e^{-\int d\hsigma V^\dagger D V}\right)
\ket{B3}_{\rm NS(RR)},
\label{inst final product}
\end{equation}
which is equal to the expression
(\ref{boundary state in the picture b})
where the gauge configuration on the $N D3$-branes is given by
\begin{equation}
 A_\mu = -V^\dagger \del_\mu V.
  \label{inst gauge field}
\end{equation}
Using the properties (\ref{zero mode condition}) and
(\ref{unitary condition 2}),
the field strength can be written as
\begin{align}
F_{\mu\nu} = \frac{1}{2}V^\dagger (\del_\mu T)
 (T^\dagger T)^{-1} (\del_\nu T^\dagger) V.
 \label{field strength pre}
\end{align}
Note that this expression of the field strength is correct for
a general tachyon profile (\ref{ADHM data}). 
Under the assumption (\ref{ADHM constraint}), 
the gauge field (\ref{inst gauge field}) is
exactly the same thing that is obtained in the ADHM construction.
In fact, using the relations (\ref{inst zero mode condition}),
(\ref{inst unitary condition 1}) and (\ref{inst unitary condition 2}),
the field strength (\ref{field strength pre}) becomes 
\begin{align}
 F_{\mu\nu} &= \frac{i}{2}\eta_{\mu\nu}^{(+)i}
 v^\dagger \left(
 {\mu_0}^{-1}\otimes\tau_i
 \right)v,  
 \label{instanton field strength}
\end{align}
where we have decomposed $V$ as 
\begin{equation}
V = \left(
\begin{matrix}
\ u \ \\
\ v \
 \end{matrix}
    \right), 
\end{equation}
with 
${N}\times{N}$ and ${2k}\times{N}$ matrices $u$ and $v$, respectively. 
{}From the self-dual nature of the 't Hooft's symbol
(\ref{self-duality of tHooft symbol}),
the field strength is obviously self-dual.
We note that it originates only from the behavior of $2\times 2$ part 
as $\del_\mu T\sim \sigma^\mu$ and $T^\dagger T \sim {\mathbf 1}_{2}$.

We here emphasize that the self-duality 
is independent of the form of low energy effective 
action of $N D3$-branes.
Namely, if we assume the BPS condition (ADHM equation) 
in the low energy theory on $k D(-1)$-branes, then,
the resulting gauge configuration after the unitary rotation 
is automatically self-dual.
Note that the limit $\alpha' \rightarrow 0$ with fixed $g_s \alpha'^{-2}$ is 
valid in the picture (a), while 
the $U(N)$ Yang-Mills theory never be valid in this limit but
(the suitable $U(N)$ generalization of) the Dirac-Born-Infeld action 
with higher derivative corrections be.
However, since this configuration is always sitting at the minimum of the 
tachyon potential irrespective of the gauge transformation,
we conclude that the self-dual field strength also
minimize the full low energy effective action of the $N D3$-branes.

Next, we discuss for the $D(-1)$-brane charge (RR $0$-form charge) 
of this boundary state (\ref{inst final product}).
Since the Chern-Simons term obtained from (\ref{inst final product}) 
is given by
\begin{equation}
S_{\rm CS} = T_3 \int_{\R^4} C \wedge \Tr_{\!\!N} 
e^{2\pi {F}}, 
\end{equation}
this state couples to the RR $0$-form $C^{(0)}$ through the 
second Chern class $\int \Tr_{\!\!N} (F\wedge F)$ for the Chan-Paton bundle.
Usually, in the context of the ADHM construction, 
the RR-charge carried by the self-dual field strength 
is estimated by using the Osborn's identity \cite{Corrigan:1979di,Osborn:1979bx},
\begin{equation}
 \Tr_{\!\!N}\left(F_{\mu\nu}F^{\mu\nu}\right) =
  -\del_\mu^2 \del_\nu^2 \log \det {\mu_0}^{-1}, 
\end{equation}
and the fact that ${\mu_0}^{-1}$ asymptotically behaves as $|x|^{-2}$. 
Then, the instanton number can be calculated as
\begin{align}
 \nu[A_\mu] &= -\frac{1}{16\pi^2}\int d^4x
 \Tr_{\!\!N}\left(F^{\mu\nu}*F_{\mu\nu}\right) \nn \\
 &= \frac{1}{16\pi^2}\int d^4x \del_\mu^2 \del_\nu^2 \log \det {\mu_0}^{-1}
 \nn \\
 &= \frac{1}{16\pi^2}\int_{S^3} d\Omega^\mu \del_\mu
 \del_\nu^2 \Tr_{\!\!k} \log {\mu_0}^{-1}
 \nn \\
 &= k, 
\end{align}
where $d\Omega^\mu$ is the area element of 
$S^3$ at infinity.
Therefore, the gauge field (\ref{gauge field}) is an
element of the $k$-instanton sector of the gauge field.
Here note that the self-duality of the field strength 
is used to obtain this result.
However, as its topological nature, the RR-charge is 
independent of the self-duality.
It can be seen directly by evaluating the asymptotic behavior 
of the gauge transformation $U_1=(V,W)$ in (\ref{U_1}) at $|x|\to\infty$:
\begin{equation}
 U_1\sim \left(
\begin{matrix}
g(x) & 0 \\
 0 & -\frac{\sigma_\mu x^\mu}{|x|}\otimes {\mathbf 1}_k
\end{matrix}
	 \right). 
\end{equation}
Here the asymptotic behavior of the Dirac zero modes $V$ is 
denoted by $g(x)\in U(N)$.
This behavior is common for any tachyon profile (\ref{ADHM data}), 
that is, independent of the self-duality.
{}From this, because the gauge field (\ref{gauge field}) has the form
$A_\mu \sim g(x)^{-1}\del_\mu g(x)$ at $|x|\to\infty$,
the second Chern class carried by the gauge connection
is also evaluated through the Chern-Simon $3$-from 
$K=\Tr(AdA+\frac{2}{3}A^3)$ at infinity, which says that 
$g(x)$ carries the mapping class $k \in \pi_3(U(N))$.
On the other hand, each component of the lower right block is 
a map from $S^3$ at infinity to $SU(2)$ with wrapping number 
$-1 \in \pi_3(SU(2))$
so that the $k$ times of this 
carries the information on the second Chern class $-k$. 
Therefore, the net second Chern class carried by $U_1$ is zero.
As we noted in the section 3, this gauge transformation 
(\ref{gauge transformation}) is a {\it large} gauge 
transformation but it does not change the topological sector of the 
total system. 
The $D(-1)$-charge originally carried by the 
$k D(-1)$-branes are now contained in the pure gauge but topologically 
non-trivial gauge field.

As a final remark, we comment on the anti-self-dual case.
The $k \overline{D(-1)}$ configuration is obtained by replacing 
$\sigma^\mu$ by $\bar{\sigma}^\mu$ in the tachyon profile (\ref{ADHM data}) 
of the picture (a). 
It leads to the field strength (\ref{instanton field strength}) in the 
picture (b) with 
replacing $\eta_{\mu\nu}^{(+)i}$ by $\eta_{\mu\nu}^{(-)i}$, which 
is anti-self-dual and has RR $0$-form charge $-k$.

\subsection{Small instanton singularities}

\begin{figure}[t]
\begin{center}
\includegraphics[scale=0.55]{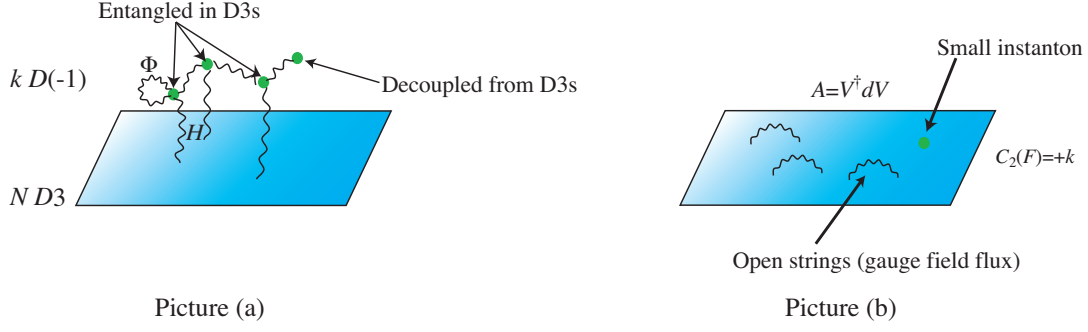}
\end{center}
\caption{\small
 A sketch of a small instanton singularity in our point of view.
 In the picture (a), some $D(-1)$-branes are decoupled from
 $D3$-branes. Correspondingly, we can define the gauge
 transformation only for a part of the Chan-Paton indices entangled
 in $D3$-branes.
 Then there are individual $D(-1)$-branes even in the picture (b),
 which makes the gauge configuration on $D3$-branes diverge. 
 }
\label{small instanton}
\end{figure} 

A small instanton singularity is a conical singularity in the instanton
moduli space of k-instantons.
Correspondingly, some of the single instantons shrink to zero size, 
thus, it is not allowed as a regular classical solution of the $4D$
gauge theory.  
In our view of the ADHM construction as a gauge transformation, 
this singularity is also understood as follows. 
For simplicity, let us set $H=0$ in the tachyon profile (\ref{ADHM data}).
Then, the tachyon profile satisfying the ADHM equation 
(\ref{ADHM constraint}) gives rise to 
\begin{align}
 T^\dagger T &=u^2 \mu_0 \otimes {\mathbf 1}_{2}
 = u^2 |\Phi-\bX|^2 \otimes {\mathbf 1}_{2}.
\end{align}
A typical example is a diagonal one 
$\Phi^\mu={\rm diag}(\phi^\mu_1,\cdots,\phi^\mu_k)$.
Because $T^\dagger T$ have zeros, we cannot 
apply the gauge transformation of the type (\ref{U_1}).
It is rather better to say that we do not need 
such a transformation since the tachyon potential is already 
block diagonal.
It leads that, in the $(N+2k)D3$/$2k\overline{D3}$ system,
$2k$ pairs of $D3/\overline{D3}$-branes always represents 
$k D(-1)$-branes while $N D3$-branes is topologically trivial.
This means simply that descriptions of $N D3$-branes 
corresponding to the picture (b)  
do not exist.
A crucial fact here is that for $H=0$ the $N D3$-branes decouple from 
$2k$ pairs of $D3/\overline{D3}$-branes.
Then it is easily generalized for the case with 
some sub-pairs of $D3/\overline{D3}$-branes decoupled, 
say $k_1$, that corresponds to the matrix $H^\dagger H$ has 
$k_1$ zero eigenvalues (see Fig. \ref{small instanton}). 
In that case, we can rotate the Chan-Paton indices 
for $k_2=k-k_1$ pairs by the gauge transformation 
and we have smooth $k_2$-instanton 
on the $N D3$-branes, while $k_1 \, D(-1)$-branes remain point-like.
This is consistent with the understanding in the picture (a) at low energy,
where the small instanton singularity is the sign for 
the connection with the Coulomb branch.

It is instructive to see the zero size limit of the instanton 
configuration in an explicit example.
Let us consider the $N=2$, $k=1$ case \cite{Belavin:1975fg}. 
In this case, the tachyon profile satisfying the ADHM equation and the 
corresponding $\mu_0$ 
are given by
\begin{equation}
 T(\bX)=u\left(
\begin{array}{c}
\rho \otimes {\mathbf 1}_{2} \\
 (\phi^\mu - \bX^\mu) \otimes\sigma_\mu
\end{array} 
    \right),
\qquad 
\mu_0= |\phi-\bX|^2 + \rho^2,
\label{BPST tachyon}
\end{equation}
where $\phi^\mu$ and $\rho$ correspond to the position and the size moduli 
of the instanton, respectively. 
{}For $\rho\ne 0$, the gauge transformation in $U(N+2k)$ 
is well-defined as
\begin{equation}
 U_1= \frac{1}{\sqrt{\mu_0}}\left(
\begin{array}{cc}
(\phi^\mu - \bX^\mu)\otimes\bar{\sigma}_\mu &
\rho \otimes {\mathbf 1}_{2} \\
-\rho \otimes {\mathbf 1}_{2} &
 (\phi^\mu - \bX^\mu)\otimes\sigma_\mu\\
\end{array} 
    \right),
\label{BPST U_1}
\end{equation}
which gives rise to the BPST instanton solution \cite{Belavin:1975fg}, 
whose field strength is
\begin{align}
 F_{\mu\nu}=\frac{2i \rho^2  \eta_{\mu\nu}^{(+)i}\tau_i}{(|x-\phi|^2+\rho^2)^2}.
 \label{BPST}
\end{align}
When the instanton size $\rho$ shrink to zero, (\ref{BPST}) approaches 
the $\delta$-function peaked at $x=\phi$. 
As mentioned above, this limit does not exist because $\mu_0$ has zero 
at a point $x=\phi$. 
This is seen in the gauge transformation (\ref{BPST U_1}):
it is not unitary so ill-defined at $x=\phi$, 
and at $x\ne\phi$ it is well-defined but valued in $U(N)\times U(2k)$, 
which means that $N D3$-branes and $k D(-1)$-branes are decoupled.
Therefore, the small instanton is realized only 
as a $D(-1)$-brane.
It is interesting that both of the expressions (\ref{BPST}) and
(\ref{ADHM tachyon}) are 
reminiscent of two expressions of the $\delta$-function,
\begin{align}
 \delta^{(4)}(x-\phi)=
\lim_{\rho\rightarrow 0}\frac{\rho^2}{(|x-\phi|^2+\rho^2)^2}
=\lim_{u\rightarrow \infty}\frac{u^4}{\pi^2}e^{-u^2 |x-\phi|^2}.
\label{delta-function}
\end{align}
Although the two parameters $\rho$ and $u$ play the completely 
different role, it might be interesting to investigate relations between
them (see the section 4.5).

\subsection{Equivalence beyond BPS}
\label{general ADHM}

So far, we have assumed that the tachyon profile (\ref{ADHM data}) 
satisfies the ADHM equation (\ref{ADHM constraint}) 
and showed that the obtained gauge field 
(\ref{gauge field}) is self-dual (\ref{instanton field strength}).
However, it is apparent from the derivation that,
as far as we have kept the assumption that $T^\dagger T$ is strictly 
positive definite, 
the gauge transformation (\ref{U(X)}) can be defined and 
(\ref{gauge field}) still gives a non-trivial
gauge field configuration on the $N D3$-branes 
even if one does not assume the ADHM condition. 
In other words, the equivalence between (a) and (b) holds independent
of the BPS condition.

To see what happen if we does not assume the ADHM condition in the picture (a), 
let us first estimate the field strength on the $N D3$-branes in the picture (b),
corresponding to (\ref{gauge field}) without the ADHM condition but assuming 
$T^\dagger T$ is positive definite. 
Recall that the expression of the field strength
(\ref{field strength pre}) is applicable to a general tachyon profile. 
Since $(T^\dagger T)^{-1}$ is a $2k\times 2k$ matrix in general, 
it is expanded with respect to the Pauli matrices $\sigma_\mu=(1,-i\tau_i)$ as 
\begin{equation}
 (T^\dagger T)^{-1} \equiv \sum_{\mu=0}^3
  f_\mu \otimes \sigma_\mu,
\end{equation}
where $\{{}^{\exists}\!f_\mu\}$ is $k\times k$ matrices written 
by $\{\mu_0,\, \mu_i\}$.
Then, since $\del_\mu T$ is unchanged, the field strength can be written 
as 
\begin{align}
 F_{\mu\nu} &= \frac{i}{2} \eta_{\mu\nu}^{(+)i}
 v^\dagger \left(f_0 \otimes \tau_i\right) v
 + \frac{i}{2}\sum_{i=1}^3 
 \eta_{\mu\nu}^{(-)i} v^\dagger 
 \left(f_i \otimes {\mathbf 1}_2\right) v,
 \label{field strength}
\end{align}
where we have used the identity, 
\begin{align}
 \frac{1}{2}\left(\sigma_\mu \sigma_i \bar{\sigma}_\nu -
 \sigma_\nu \sigma_i \bar{\sigma}_\mu \right) &=
 i\eta_{\mu\nu}^{(-)i}\, {\mathbf 1}_{2},  \qquad (i=1,2,3)
\end{align}
in addition to (\ref{identity-1}). 
Note that $v$ in (\ref{field strength}) is different from that given in 
(\ref{instanton field strength}) and $f_0\ne (\mu_0)^{-1}$,
but of course if we impose the ADHM condition (\ref{ADHM constraint}),
it reproduces the self-dual field strength 
(\ref{instanton field strength}). 
In the present case, an anti-self-dual component  
(the second terms of (\ref{field strength})) corresponding to
$f_i$ is introduced in the field strength.
In the low energy Yang-Mills theory, such a deviation 
from the instanton solution increases the action (mass of the $D3$-branes). 
{}From the view point of the string theory, this corresponds to
the deviation from the BPS condition, 
since the RR-charges are preserved 
even if the tachyon configuration does not satisfy
the ADHM condition. 
Since the ADHM equation is originally 
the BPS condition for the $k D(-1)$-brane effective theory and 
its breaking raises the mass, this simply means that 
the supersymmetry breaking occurs as the same manner in both pictures.

This is a direct consequence of the fact 
that the picture (a) and (b) are equivalent in the 
full string theory even if the BPS condition is not satisfied,
since it is simply due to the gauge equivalence 
in the system of $(N+2k) D3$-branes and $2k \overline{D3}$-branes
under the gauge symmetry $U(N+2k)\times U(2k)$.
We here summarize our claim. 
The system of $N D3$/$k D(-1)$ bound state in the picture (a) and the 
system of $N D3$-branes with fluxes in the picture (b)
are gauge equivalent with each other. 
Each picture corresponds to a particular gauge:
\begin{align*}
\text{(a)}  \left\{
\begin{array}{ll}
\text{background:} & k D(-1) + N D3 \\
\text{fluctuations:} & (H,\Phi^\mu) 
\end{array}
\right.\,\,\
\text{(b)} \left\{
\begin{array}{ll}
\text{background:} & N D3 \\
\text{fluctuations:} & A_\mu=-V^\dagger \del_\mu V .
\end{array}
\right.
\end{align*}
Here the background means a particular boundary state
defining a conformal field theory on the disk,
and fluctuations are boundary interactions acting on it.
The equivalence says that
the backgrounds of both pictures are 
apparently different but the 
total systems with adding fluctuations are the same. 
It is because both the background and the fluctuations are
determined by the configurations for tachyon and gauge fields 
(or simply the boundary interaction matrix $\bM$) on 
the $(N+2k) D3$/$2k \overline{D3}$-system.
In particular, it gives the correspondence between 
DN- and DD-strings $\{H,\Phi^\mu\}$ 
in the picture (a) and NN-strings $-V^\dagger \del_\mu V$ in the picture (b).
This equivalence is quite general in the sense that 
it is independent of the BPS condition and is valid at all order in $\alpha'$.
If we additionally impose the BPS condition in both sides, 
it gives the ADHM construction in the low energy limit. 
As mentioned, imposing the ADHM condition on $\{H,\Phi^\mu\}$ corresponds 
to restricting the field strength be self-dual.
Moreover, the fluctuations $\{H,\Phi^\mu\}$ rotated by the 
$U(k)$ symmetry gives the same $A_\mu$,
and the gauge transformation itself has ambiguity $U(N)$,
which is seen as the (topologically trivial) gauge symmetry in the picture (b).
Therefore, we have obtained the map, 
\begin{equation}
 \frac{\{(H,\Phi^\mu)|\mbox{ADHM eqs.}\}} {U(k)}
 \rightarrow
 \frac{\{-V^\dagger \del_\mu V|\mbox{self-dual eqs.}\}} {U(N)}.
 \label{moduli map}
\end{equation}
If we can replace $-V^\dagger \del_\mu V$ by an arbitrary gauge field $A_\mu$, 
this map means the isomorphism between the ADHM moduli space and the 
instanton moduli space%
\footnote{ 
To show the isomorphism, we need a gauge transformation starting from the
picture (b).
We can also prove it through the so-called inverse ADHM construction 
\cite{Hashimoto:2005qh}.}.

In the above claim, there is a remaining task to show that $2k$ pairs of 
$D3$/$\overline{D3}$-branes disappear even if the system is non-BPS.
There also arises a natural question, how one can decompose a 
given field configuration
into the background and fluctuations, 
or whether this decomposition is unique.
From the $(N+2k) D3$/$2k \overline{D3}$ point of view it is abuse question
and it is a matter of the physical interpretation.
It is also related to the definition of the moduli space 
in the boundary state setting.
In the low energy effective theory, 
in which a background determines the base space (worldvolume) and fluctuations 
become fields on it,
the notion of the moduli space is well-defined.
Here let us look at the Fig.~\ref{strategy} again and recall 
that our argument relies on two independent notions,
\begin{enumerate}
\item gauge transformation $U(N+2k)\times U(2k)$,
\item tachyon condensation $u\rightarrow\infty$,
\end{enumerate}
where the first item corresponds to the horizontal 
line in the Fig.~\ref{strategy} and 
the second represents two vertical lines.
We have so far mainly concentrated ourselves on the first item.
In general, any two configurations of 
the $(N+2k) D3$/$2k \overline{D3}$-system are gauge equivalent
with each other as far as there exists a gauge transformation 
(which is not restricted to a particular form (\ref{U(X)})) 
connecting them. 
On the other hand, our questions are concerned to the second item,
so we next discuss the second item in more detail.

Let us examine the boundary interaction 
in more detail for the tachyon profile (\ref{ADHM data}).
First, for simplicity, we set $H=0$ in the picture (a), 
corresponding to small instantons.
In this case, $N D3$-branes are decoupled 
and the boundary interaction $e^{-S_b}$ in the NSNS-sector 
is given as
\begin{align}
 e^{-S_b} \ket{B3}
= N \ket{B3}+ e^{-\widetilde{S}_b} \ket{B3}, 
\end{align}
where the first term is the decoupled $N D3$-branes.
We concentrate on the second term,
which is $2k D3$/$\overline{D3}$-branes with the 
boundary interaction given as
\begin{align}
 e^{-\widetilde{S}_b} \ket{B3}&= {\rm Tr}_{\!4k}
 \hP \exp\left\{\int d\hsigma 
 u(\Phi^\mu-\bX^\mu)\Gamma_\mu 
 \right\} \ket{B3} \nn \\
 &= {\rm Tr}_{\!4k} P \exp\left\{\int d\sigma \left(
 -{u^2}|\Phi^\mu-X^\mu|^2 {\mathbf 1}_{4}
 -{u^2}[\Phi^\mu,\Phi^\nu] \Gamma_{\mu\nu}
 +u\Psi^\mu \Gamma_\mu
\right)\right\} \ket{B3}\nn \\
 &= \int[d\zeta_\mu] \,{\Tr}_{\!k} P \,\, e^{\int d\sigma\left(
 \frac{1}{4}{\zeta}_\mu \del_\sigma{\zeta}_\mu
  -{u^2}|\Phi^\mu-X^\mu|^2
  -{u^2}[\Phi^\mu,\Phi^\nu] \zeta_{\mu\nu}
  +u\Psi^\mu \zeta_\mu
 \right)} \ket{B3} \nn \\
 &= \int[d\zeta_\mu][dX_\mu][d\Psi_\mu]\, {\Tr}_{\!k} P \,\,
  e^{\int d\sigma\left(
  \frac{1}{4u^2}\zeta_\mu \del_\sigma {\zeta}_\mu
  -{u^2}|\Phi^\mu-X^\mu|^2
  -[\Phi^\mu,\Phi^\nu] \zeta_{\mu\nu}
  +\Psi^\mu \zeta_\mu
 \right)}\ket{\bX^\mu},
 \label{dependence on u}
\end{align}
where
\begin{equation}
 \Gamma_\mu \equiv \left(
\begin{matrix}
 0 & \sigma_\mu \\
 \bar{\sigma}_\mu & 0
 \end{matrix}
\right), 
\end{equation}
is the $4$-dimensional Dirac gamma matrices
and
$\Gamma_{\mu\nu}\equiv\frac{1}{2}\left[\Gamma_\mu,\Gamma_\nu\right]$.
{}From the second line to the third line,
we have replaced $\Gamma_\mu$ by the boundary fermions $\zeta_\mu (\sigma)$ 
with the anti-periodic boundary
condition 
and replaced the trace over the Clifford algebra
by the functional integral \cite{Kraus:2000nj}.
{}From the third line to the last line, 
we have rescaled $\zeta_\mu\to \zeta_\mu/u$
which does not change the measure $[d\zeta_\mu]$ 
of the anti-periodic boundary
fermions.
In the language of the decomposition (\ref{expansion of TT}),
the second term of (\ref{dependence on u}), $|\Phi^\mu-X^\mu|^2$
is equal to $\mu_0$ and the third term, $[\Phi^\mu,\Phi^\nu]\zeta_{\mu\nu}$ is 
a linear combination of $\mu_i$. 
By performing the functional integral and taking the limit
$u\rightarrow \infty$, 
it reduces to (\ref{scalar}),
\begin{equation}
 {\rm Tr}_k \, \hP e^{-i\int d\hsigma \Phi^\mu
  \bP_\mu} \ket{\bX^\mu=0}
 ={\rm Tr}_k \, P 
 e^{\int d\sigma \left(-i\Phi^\mu P_\mu
 -[\Phi^\mu,\Phi^\nu] \Pi_\mu \Pi_\nu \right)} \ket{\bX^\mu=0}. 
 \label{scalar fluctuation}
\end{equation}
 
The expression (\ref{dependence on u}) represents a system
(\ref{scalar fluctuation}) 
of $k D(-1)$-brane with scalar fields 
from the $2k D3$/$\overline{D3}$-branes point of view. 
Roughly speaking, $\mu_0$ ($\mu_i$) in (\ref{dependence on u}) 
corresponds to the first (second) term of (\ref{scalar fluctuation}). 
By comparing (\ref{scalar fluctuation}) with  (\ref{dependence on u}), 
we can read off more precise information on the tachyon condensation.
When $\Phi^\mu=0$, they represent coincident $k D(-1)$-branes, 
which define a background $k \ket{\bX^\mu=0}$
and $\Phi^\mu$ are fluctuations around this background.
In the low energy effective theory of $k D(-1)$-branes,
$\Phi^\mu$ become scalar fields on the $0$-dimensional worldvolume,
but $\mu_0$ and $\mu_i$ play the different role.
In fact, $\mu_i$ term gives rise to the potential term 
$\Tr_k |[\Phi^\mu,\Phi^\nu]|^2$ and hence raises the mass,
while $\mu_0$ term works as the (generalized) pull-back of 
embedding the worldvolume into the space-time.
{}From the worldsheet point of view, the insertion of 
$\int d\hsigma \Phi^\mu \bP_\mu$ in (\ref{scalar fluctuation}) 
gives a marginal deformation of the disk with the Dirichlet boundary condition,
because its conformal dimension is $1$.
As far as they are treated perturbatively,
such fluctuations do not change the background,
but if arbitrary numbers of insertions are allowed and 
if they are mutually commuting,
such marginal perturbation could change the background \cite{Recknagel:1998ih}.
For example, if 
$\Phi^\mu={\rm diag.}(\phi^\mu_1,\cdots,\phi^\mu_k)$ are diagonal,
$\mu_i=0$ and  
(\ref{scalar fluctuation}) becomes
\begin{equation}
 \ket{X^\mu=\phi_1,\Psi^\mu=0}+\cdots +\ket{X^\mu=\phi_k,\Psi^\mu=0},
\end{equation}
which represents $k D(-1)$-branes
located at $x^\mu=\phi^\mu_1,\cdots,\phi^\mu_k$.
This is an example of the exactly marginal deformation from 
coincident $k D(-1)$-branes.
Here $4k$ parameters of the location span the moduli space.
This means that in terms of (\ref{dependence on u}), 
the principal part of the tachyon condensation is $\mu_0$ with $\mu_i=0$.
Namely, the insertions of $\mu_0$ 
with $\mu_i=0$ represent a relevant deformation 
of the worldsheet,
in which the disk with the Neumann boundary condition is deformed 
to that with the Dirichlet boundary condition,
and the possible choice of $\mu_0$ span the same moduli space.
This is also seen by the $u$-dependence in (\ref{dependence on u}): 
only $\mu_0$ contributes to the tachyon condensation $u\rightarrow\infty$,
while $\mu_i$ are $u$-independent and hence considered to be 
fluctuations around the condensed background.
Physically, it reflects the fact that the $D3$/$\overline{D3}$-system 
is strongly bounded but the resultant $D(-1)$-branes are 
marginally bounded with each other.

Some remarks are in order.
First, the fact that $\mu_0$ has always $k$ zeros at some points
(i.e., $T^\dagger T$ with $\mu_i=0$ has $2k$ zeros), 
means the resultant defects are small instantons. 
Second, if we regard whole $T^\dagger T$ rather than $\mu_0$ as the principal part, 
or equivalently, if we take the basis diagonalizing $T$ rather than $\mu_0$, 
it never leads to the picture (a) except for the BPS case 
but leads to another picture%
\footnote{This is the treatment of the tachyon condensation in 
\cite{Hashimoto:2005qh}.},
because $\mu_i$ terms necessarily split the eigenvalues of $T^\dagger T$
and $T^\dagger T$ has only $k$ zeros. 
It also breaks the structure of the fermion 
$\Psi^\mu$ terms, which is also important for defining the background.

The situation is in principle the same when we take into account of $H$. 
The tachyon profile (\ref{k D(-1) solution}) 
gives $N D3$-branes and $k D(-1)$-branes without 
open strings, which defines a background in the picture (a). 
By turning on the massless open strings $\{H,\Phi^\mu\}$ as perturbations,
they still represent the same system with fluctuations.
However, 
if we include $\{H,\Phi^\mu\}$ which satisfy the ADHM equations $\mu_i=0$ 
as background, the condensation of such term
represents $N D3$/$k D(-1)$ bound state, 
which is a exactly marginal deformation from (\ref{k D(-1) solution}).
All possible choices are equally regarded as the backgrounds 
for the string theory, that is,
they span the moduli space of the $N D3$/$k D(-1)$ bound state.
This is the stringy explanation of the ADHM moduli space.
On the other hand, $\{H,\Phi^\mu\}$ with non-zero $\mu_i$ 
are considered to be fluctuations from this bound state.
Therefore, within the picture (a) the decomposition 
into the background and the fluctuations 
itself is not unique, but for any choice the total system is the same.

Under the gauge transformation, this structure is mapped to 
the picture (b).
Treating $\mu_0$ with $\mu_i=0$ as the principal part of the tachyon condensation
is the same as taking $P_0=\mu_0$ in (\ref{separation formula}) in the section 3.
It is exactly the ADHM construction argued in the section 4.3.
If $\mu_0$ is strictly positive definite (no small instantons), 
$2k$ pairs of $D3$/$\overline{D3}$-branes are always annihilated 
into the vacuum for any choices of $\mu_0$ and 
a self-dual gauge field appears on the remaining $N D3$-branes.
In other words, possible backgrounds of the picture (b) are 
$N D3$-branes with self-dual field strength, which depends on moduli
parameters. 
On the other hand, the effect of fluctuations are contained not only 
in $\mu_i$ but also in 
the pure gauge part $\bA$ in (\ref{separation formula}),
since the gauge transformation is defined not by $\mu_0$ but by whole $T$.
Note that the fate of $2k$ pairs of $D3$/$\overline{D3}$-branes is 
unchanged by including the $\mu_i$ term because they are treated 
perturbatively and becomes hidden into the vacuum.
Therefore, fluctuations are seen only as the deviations of the gauge field 
from the instanton solution.
This explain the correspondence with the deviation from the ADHM equations 
in the picture (a), as we demonstrated 
at the beginning of this subsection.

Along this line, we can further extend our analysis.
Other kinds of open string excitations can be incorporated as fluctuations 
in both sides.
An rather trivial example is turning on 
the transverse scaler fields $\Phi^i$ on the $N D3$-branes.
In this case, if the system is BPS, they define the new backgrounds 
with $U(N)$ gauge symmetry breaking as seen from the low energy effective theory.
Another example is turning on the scalar fields on the $2k$ pairs of 
$D3$/$\overline{D3}$-branes, which corresponds to the Coulomb branch 
in the picture (a).
More interesting possibility is an insertion of massive excitations.
If the tachyon condensation parameter $u$ is quite large but finite,
such a non-infinity effect can be regarded as a massive excitation 
around the $u=\infty$ background \cite{Asakawa:2005vb}, 
that is, $k D(-1)$-branes in the 
picture (a).
Corresponding boundary state is not completely localized 
but has a size $\sim 1/u$, or equivalently, it possesses
a mixed boundary 
condition on the disk depending on $u$.
Such a system would correspond to a gauge configuration 
in the picture (b), where the defect has minimal size $\sim 1/u$,
analogous to the noncommutative instanton, where
the existence of B-field changes the boundary condition and 
possesses the minimal size $\sim \theta$.
In this respect, the equation (\ref{delta-function}) could 
have another meaning.
The mechanism presented in the section 3 is also applicable 
to the $3$ charge system or more complicated D-brane bound states.
For example, when scalar fields $3$ of $\Phi^\mu$ satisfy the 
$SU(2)$ Lie algebra, they couple to the RR $2$-form of constant 
field strength known as the Myers effect \cite{Myers:1999ps},
then it is better to think that there are also spherical $D1$-branes 
as the background.
In this case, the background of the pictures of (a) and (b) are no
longer good backgrounds and there should be a more suitable background
which is achieved by a relevant perturbation from them. 
In any case, however, it is still true that 
there are several pictures related among them 
by the gauge transformation in the $D3$/$\overline{D3}$-system.

\section{Generalizations}

In the previous section, we have concentrated on 
the construction of instantons which is the codimension 4
bound state of $D(-1)$-branes within $D3$-branes. 
One might naively expect that we can repeat completely the same arguments in the
previous sections and we should obtain in general the bound states 
with even codimensions $D(-1)$-$D(2n-1)$
after the tachyon condensation.
The equivalence between the $D$-brane bound state
and gauge field on $D$-branes  is still true as we discussed in the section 4.5.
However, the final products would not be stable since the bound states are not
BPS and break the supersymmetry generally.
Let us see here the case of
codimension 2 (vortex) and codimensions greater than 4 in detail.

\subsection{2 dimensional vortex}

We first consider the codimension 2 case, namely $D(-1)$-$D1$ case as vortices.
In this case, we need the system of $(N\!+\!k) D1$-branes and 
$k \overline{D1}$-branes, and 
the chiral decomposed gamma matrices are simply $\gamma_0=1$ and
$\gamma_1=i$.
So the tachyon field in the picture (a) 
is a holomorphic function of the complex coordinate $z=x^0+ix^1$
valued in $(N\!+\!k)\times k$ matrices
\be
T(z)=u\left(
\begin{array}{c}
H\\
\Phi - z
\end{array}
\right),
\ee
where $\Phi$ is a $k\times k$ hermitian matrix and $H$ is a $N\times k$ matrix.

Using the holomorphy of the tachyon $T(z)$, we easily find the 2d analog of the
Osborn's identity\footnote{In contrast with the 4d instanton case,
this identity always satisfies without any self-dual like condition.}
\cite{Corrigan:1979di, Osborn:1979bx}, 
\be
\Tr F = \del\overline{\del}\log \det T^\dag T,
\label{2d Osborn}
\ee
where we defined $F\equiv \epsilon_{\mu\nu}F^{\mu\nu}$.
Since $T^\dag T$ behaves asymptotically as $\det T^\dag T \sim u^2|z|^{2k}$
in the $|z|\rightarrow\infty$ limit, we obtain the vorticity by
\be
\frac{1}{2\pi i}\int d^2 z \, \Tr F = \frac{k}{2\pi i}\oint \frac{dz}{z} = k.  
\ee

As well-known, the codimension 2 bound state is not stable and the localized
(finite size) vortex configuration is impossible to be the BPS state.
The energy of the vortex is minimized
at the large size limit and the $D(-1)$-brane (vortex) should
dissolves in the $D1$-brane.
 (See Sec. 13.6 in \cite{Polchinski:1998rr}, for example.) 
To see this, it is sufficient to estimate the Yang-Mills energy
for $U(1)$ one vortex ($N=k=1$).
The tachyon field is given by
\be
T(z)=\left(
\begin{array}{cc}
h\\
\varphi-z
\end{array}
\right),
\ee
where $\varphi$ and $h$ are complex parameters which represents the position and
size moduli of
the vortex, respectively. Indeed, from (\ref{2d Osborn}), we find
\be
F = \del\overline{\del}\log(|\varphi-z|^2+|h|^2)=\frac{|h|^2}{(|\varphi-z|^2+|h|^2)^2},
\ee
whose distribution gives the meanings of the moduli parameters. From this
$U(1)$ field strength, we can see
\be
\frac{1}{2\pi i}\int d^2 z F = 1,
\ee
independently of $|h|$ as expected.
Contrarily, the Yang-Mills energy,
\be
\begin{split}
E_{\rm YM} &= \frac{1}{4g_{\rm YM}^2}\int d^2 z\, F^2\\
&=\frac{\pi}{12g_{\rm YM}^2}\frac{1}{|h|^2},
\end{split}
\ee
depends on the size moduli $|h|$ and
diverges in the limit of $|h|\rightarrow 0$. This means that the minimum of
the energy corresponds to the $|h|\rightarrow \infty$ limit. So
the $D(-1)$-brane is not localized on $D1$-brane any longer.

From our viewpoint, it is still true that 
the $D(-1)$-$D1$ bound state in the picture (a) and 
$D1$-branes with vortices in the picture (b)
are gauge equivalent with each other.
However, both pictures have instability as the same manner.
It is seen in the picture (a) that DN-strings $H$ have tachyonic modes, 
while in the picture (b) the same $H$ possesses instability of 
field strength $F$ as described above. 
Therefore, in this case, these pictures are not good descriptions.
It is not a matter of the gauge equivalence but that of the tachyon condensation
in the terminology of the section 4.5.
If we are interested in the stable BPS state, we should consider 
other situation like the presence of B-fields or other massless fields, 
where there are some stabilization effects.
Once we have found such a BPS D-brane system, the gauge equivalence would give 
the construction of corresponding soliton.

\subsection{Instantons in dimension greater than four}

We next consider the even codimension greater than $4$ case.
As in the codimension $2$ case, $D(-1)$-$D5$ or 
$D(-1)$-$D7$ bound states are not BPS so that 
the naive extension of the $4$-dimensional instantons does not hold
in the higher dimensional case.
However, as opposed to the codimension $2$ case, it is possible 
to find BPS state within the Yang-Mills theory,
that is, no additional fields are needed in the low energy effective 
theory of the picture (b).

In order to obtain the stable BPS states, 
we need to impose an extended ``self-duality'' condition
for the gauge field in even dimensions greater than four
\cite{Corrigan:1982th, Ward:1983zm}. 
The extended
``self-dual'' conditions are expressed as
\be
\frac{1}{2}T_{\mu\nu\rho\sigma}F^{\rho\sigma}
=\lambda F_{\mu\nu},
\label{extended self-dual}
\ee
where $T_{\mu\nu\rho\sigma}$ is a totally antisymmetric tensor and
$\lambda$ is an eigenvalue.
If $F_{\mu\nu}$ satisfies the condition (\ref{extended self-dual}),
the e.o.m.~$D^\mu F_{\mu\nu}=0$ is trivially satisfied due to the Jacobi
(Bianchi) identity. 
This linealization of the e.o.m.~strongly suggests that the extended self-dual
equation is integrable and the solutions give the BPS bound states. 
In contrast with the 4 dimensional case,
the condition (\ref{extended self-dual}) can not be invariant under the
whole Lorentz group $G=SO(2n)$, but invariant under a subgroup $H\subset G$.
Therefore the representation of the 4-form tensor $T_{\mu\nu\rho\sigma}$
must contain a singlet piece under $H$ at least.

In the 6 dimensional case, $T_{\mu\nu\rho\sigma}$ belongs to ${\bf 15}$ of
$G=SO(6)$.
The non-trivial decomposition which does not reduce the 4 dimensional instanton
and contains a singlet piece can be done by choosing a subgroup 
$H=[SU(3)\times U(1)]/\Z_3$.
Under this $H$, the antisymmetric tensor $T_{\mu\nu\rho\sigma}$ and 
the field strength $F_{\mu\nu}$, which is also ${\bf 15}$ 
representation, decompose into
\be
{\bf 15} = {\bf 1}_0 + ({\bf 3}_2+\overline{\bf 3}_{-2}) + {\bf 8}_0,
\label{15 decomposition}
\ee
where the subscripts stand for the $U(1)$ charges,
and the singlet, $({\bf 3}+\overline{\bf 3})$ and octet pieces correspond to
the eigenvalues $\lambda = -2,-1,1$, respectively.
Similarly, an asymmetric product of the gamma matrices
(\ref{higher dim gamma}) decomposes into (\ref{15 decomposition}) as
\be
\Gamma_{[\mu}\Gamma_{\nu]}
=\eta^{(-2)}_{\mu\nu}
+(\eta^{(-1)a}_{\mu\nu}T_a
+\eta^{(-1)\overline{a}}_{\mu\nu}T^\dag_{\overline{a}})
+\eta^{(+1)i}_{\mu\nu}\hat{T}_i,
\ee
where $T_a$, $T_{\overline{a}}$ and $\hat{T}_i$ ($a,\overline{a}=1,\ldots,3$
 and $i=1,\ldots,8$) are
the generators in $Spin(6)$ representation and
$\eta^{(-2)}_{\mu\nu}$, $\eta^{(-1)a}_{\mu\nu}$, $\eta^{(-1)\overline{a}}_{\mu\nu}$
and $\eta^{(+1)i}_{\mu\nu}$
are extended 't Hooft tensors, which satisfy
\be
\frac{1}{2}{T_{\mu\nu}}^{\rho\sigma}\eta^{(\lambda)}_{\rho\sigma}
=\lambda\eta^{(\lambda)}_{\rho\sigma}.
\ee

Following the same ADHM construction  of instanton as the 4 dimensional case,
we expect that the field strength is proportional to 
$\gamma_{[\mu}\gamma_{\nu]}^\dag$
if $T^\dag T$ is proportional to the identity matrix 
(i.e., commute with any $\gamma_\mu$),
where $\gamma_\mu$ is the chiral parts of $\Gamma_\mu$ in
(\ref{higher dim gamma}). 
Unfortunately, however, 
$\gamma_{[\mu}\gamma_{\nu]}^\dag$ is not proportional to any of
$\eta^{(\lambda)}_{\mu\nu}$ in contrast with the 4 dimensional case.
This means that a naive construction of the 6 dimensional instanton via
the tachyon profile,
\be
T(\bX)=u\left(
\begin{array}{c}
H\\
(\Phi^\mu-\bX^\mu)\otimes \gamma_\mu
\end{array}
\right), 
\ee
does not give the self-dual 6 dimensional instanton (\ref{extended self-dual}).
It reflects the fact that the $D(-1)$-$D5$ BPS bound state
can not exist without $B$-field
\cite{Mihailescu:2000dn, Witten:2000mf, Ohta:2001dh}.
In order to obtain the 6 dimensional instanton configuration,
we need more generic tachyon profile which respects for the broken
Lorentz symmetry $[SU(3)\times U(1)]/\Z_3$, but 
this is out of issue in this paper.

Next, let us see the 8 dimensional case. $T_{\mu\nu\rho\sigma}$
belongs to ${\bf 70}={\bf 35} +\overline{\bf 35}$.
There are four possible subgroups $H$ of the Lorentz
group $G=SO(8)$, which are  
$Spin(7)$, $[SU(4)\times U(1)]/\Z_4$, $[Sp(2)\times Sp(1)]/\Z_2$
and $SO(4)\times SO(4)$.
The $SO(4)\times SO(4)$ case however is rather trivial, since
it is an intersection of the 4 dimensional instantons independently.
Under the non-trivial subgroups, the field strength $F_{\mu\nu}$ 
of ${\bf 28}$ is decomposed as
\begin{eqnarray}
{\bf 28} = {\bf 7}+{\bf 21},
& \text{ for } Spin(7),\\
{\bf 28} = {\bf 1}_0 +  ({\bf 6}_2 + {\bf 6}_{-2})+{\bf 15}_0,
& \text{ for } [SU(4)\times U(1)]/\Z_4,\\
{\bf 28} = {\bf 3} + {\bf 15} + {\bf 10},
& \text{ for } [Sp(2)\times Sp(1)]/\Z_2.
\label{Sp2 decomposition}
\end{eqnarray}
The strategy to obtain the BPS bound state from the tachyon condensation
is the same as the 6 dimensional case. So if we can 
choose a suitable tachyon profile 
under the above subgroup, we obtain the ADHM equations
and ``self-dual'' field strength 
for the 8 dimensional instantons,
but it is difficult to construct explicitly, 
except for the $[Sp(2)\times Sp(1)]/\Z_2$ case \cite{Corrigan:1984si,Ohta:2001dh}. 
In the $[Sp(2)\times Sp(1)]/\Z_2$ case, we can choose the tachyon profile as
\be
T(\bX) = u\left(
\begin{array}{cc}
I^\dag-K^\dag \bZ_4^\dag - L\bZ_3^\dag & J + K^\dag\bZ_3 -L\bZ_4\\
B_2^\dag - \bZ_2^\dag -B_4^\dag \bZ_4^\dag + B_3 \bZ_3^\dag
& -B_1+\bZ_1 +B_4^\dag \bZ_3 + B_3 \bZ_4 \\
B_1^\dag - \bZ_1^\dag -B_3^\dag \bZ_4^\dag -B_4\bZ_3^\dag
& B_2 - \bZ_2 + B_3^\dag \bZ_3 - B_4\bZ_4
\end{array}
\right),
\label{Sp2 tachyon}
\ee
where 
\be
\bZ_1\equiv \bX^2+i\bX^1,\quad
\bZ_2\equiv \bX^0+i\bX^3,\quad
\bZ_3\equiv \bX^6+i\bX^5,\quad
\bZ_4\equiv \bX^4+i\bX^7,
\ee
and
$k{\times}k$ matrices $B_i$ and
$N{\times}k$ matrices $(I^\dag,J,K^\dag,L)$ represents fluctuations.
Using this profile, we can repeat the same arguments as the 4 dimensional case
and obtain the gauge field strength of ${\bf 3}$ and the ADHM equations
of ${\bf 10}$ in (\ref{Sp2 decomposition}).

We can read off the $D$-brane bound state 
from the tachyon profile (\ref{Sp2 tachyon}).
Note that this tachyon is a $(N\!+2k)\times 2k$-matrix valued   
function on the $8$ dimensional worldvolume, 
that is, the system here is $(N\!+\!2k) D7$/$2k \overline{D7}$-system. 
Suggested by this, 
it is regarded as the $D3$-branes within $D7$-branes.
Indeed, the profile (\ref{Sp2 tachyon}) has exactly the same form as 
(\ref{ADHM data}) in the section 4, but the ADHM data are now 
functions of $\bZ_3$ and $\bZ_4$.
For any fixed $(\bZ_3, \bZ_4)$ it represents
codimension 4 defects (instantons) 
localized in the $4$-dimensional $(z_1, z_2)$-plane,
and its world volume extends along the $4$-dimensional $(z_3, z_4)$-plane.
It is nothing but the $D3$-branes with the $(z_3, z_4)$-dependent 
fluctuations (ADHM data) in the picture (a).
Of course in the picture (b), it represents the family of gauge instantons 
located in the $(z_1, z_2)$-plane,
whose moduli are $(z_3, z_4)$-dependent.
Since the size moduli diverges and the gauge configuration
becomes sparse at infinity $|z_3|,|z_4| \rightarrow \infty$,
the energy density is effectively localized in the $8$-dimensional space.
We can also see (\ref{Sp2 tachyon}) as the $(z_1, z_2)$-dependent ADHM data,
by a suitable change of basis so that (\ref{Sp2 tachyon}) has the canonical 
form with respect to $\bZ_3$ and $\bZ_4$.
From this viewpoint with fixed $(z_1, z_2)$, we can obtain other $D3'$-branes, 
which are 
localized in the $(z_3, z_4)$-plane and extending in the $(z_1, z_2)$-plane,
with $(z_1, z_2)$-dependent fluctuations.
So this means that in the picture (a) it represents
the intersecting $D3$-branes (instantons) at angles bounded to $D7$-branes
\cite{Ohta:1997fr,Papadopoulos:1997dg}, whose intersections
are non-trivially deformed by the fluctuations.

Finally we comment on interesting dual systems.
The breaking of the global Lorentz symmetry is closely
related to the number of supercharges preserved by
the $D(-1)$-$D(2n-1)$ bound state. 
If the bound states realize the ``self-dual''
condition, it preserves some Killing spinors associated with a holonomy group.
This fact also says that the BPS bound states are dual to 
an intersecting brane system or
curved space with the same number of the supersymmetry.
For example $D(-1)$-$D3$-bound state, we can find the following duality maps:
\be
\begin{CD}
D(-1)/D3 @>T>> D5/D5' @>S>> NS5/NS5' @>T>> \text{CY$_3$ (conifold)},
\end{CD}
\ee 
where $T$ and $S$ stand for T- and S-duality, respectively. The final curved Calabi-Yau
3-fold has $SU(3)$ holonomy, which preserves $1/4$ of supercharges as the same
number as the initial $D(-1)$-$D3$ bound state.
For other bound state with higher dimensional codimensions,
they are dual to 8 dimensional Joyce manifold,
Calabi-Yau 4-fold and hyper-K\"ahler manifold with $Spin(7)$, $SU(4)$ and $Sp(2)$
holonomy, which preserves $1/16$, $1/8$ and $3/16$ of supercharges, respectively.
It is interesting to consider the relation between the tachyon condensation and
the above special holonomy manifold.
It it generally difficult to apply the same dual maps as the above to $D$-$\overline{D}$
system.
The dual of the $D$-$\overline{D}$ pair decay
however would give a {\it closed} tachyon condensation
into the curved special holonomy manifolds without $D$-branes
but preserving the same number of the supersymmetry.

\section*{Acknowledgements}
The authors would like to thank
H.~Kawai,
H.~Suzuki,
T.~Tada, 
K.~Hashimoto,
and S.~Terashima
for useful discussions and valuable comments. 
KO also thanks M.~Nitta and K.~Ohashi.
This work is supported by Special Postdoctoral Researchers
Program at RIKEN.

\appendix

\section{Gauge Transformation of the Boundary Interaction}
\label{app gauge}

In this appendix, we prove that the supersymmetric path-ordered product
for an $M\!\times\!M$ hermitian matrix $\bM(\hsigma)$, 
\begin{equation}
 \widehat{W} \equiv \Tr \hP e^{\int d\hsigma \bM(\hsigma)},
  \label{super Wilson loop in appendix}
\end{equation}
is invariant under the transformation,
\begin{equation}
\bM(\hsigma) \to \bM'(\hsigma)=-U(\hsigma)^\dagger DU(\hsigma)
+ U^\dagger(\hsigma) \bM(\hsigma) U(\hsigma),
\label{gauge transformation in appendix}
\end{equation}
for an arbitrary unitary matrix $U\in U(M)$.
Here we assume that the matrix $\bM(\hsigma)$ is fermionic and
the unitary matrix $U(\hsigma)$ depends on the supercoordinate as
\begin{equation}
 U(\hsigma) \equiv U_0(\sigma) + \theta U_1(\sigma), 
\end{equation}
where $U_0(\sigma)\in U(M)$ and $U_1(\sigma)$ must satisfy
\begin{equation}
 U_0(\sigma) U_1^\dagger(\sigma) + U_1(\sigma) U_0^\dagger(\sigma)
  =0
  \label{unitary condition in appendix}
\end{equation}
in order to $U(\hsigma)$ be a unitary matrix. 
Note that this relation (\ref{unitary condition in appendix}) is
automatically satisfied if the unitary matrix is a function of $\bX$,
$U=U(\bX)$, as in the section 2.

In order to show the invariance of $\widehat{W}$ under (\ref{gauge
transformation in appendix}), we first write $\bM(\hsigma)$ as
\begin{equation}
 \bM(\hsigma) = M_0(\sigma) + \theta M_1(\sigma).
\end{equation}
Then we can easily see that $M_0$ and $M_1$ transform as
\begin{align}
 M_0 &\to M'_0 = U_0^\dagger M_0 U_0 + U_0^\dagger U_1, \\
 M_1 &\to M'_1 = -U_0^\dagger \del_\sigma U_0
 +U_0^\dagger M_1 U_0 - U_1^\dagger U_1
 -U_0^\dagger M_0 U_1 + U_1^\dagger M_0 U_0, 
\end{align}
under the transformation (\ref{gauge transformation in appendix}).
In particular, we can show that the combination
$M\equiv M_1 - M_0^2$ transforms as
\begin{equation}
 M \to M'=-U_0^\dagger \del_\sigma U_0 + U_0^\dagger M U_0,
  \label{usual gauge transformation}
\end{equation}
using the condition (\ref{unitary condition in appendix}). 
Recalling that $\widehat{W}$ can be written in the usual
form of Wilson loop as
\begin{equation}
 \widehat{W} = \Tr P e^{\int d\sigma M(\sigma)}, 
\end{equation}
we can immediately show that $\widehat{W}$ is invariant
under the transformation (\ref{usual gauge transformation}), 
which is the gauge transformation for a standard Wilson loop operator.
This means that $\widehat{W}$ is invariant under the transformation
(\ref{gauge transformation in appendix}).

\section{A Formula for the path-ordered product}
\label{app formula}

In this appendix, we evaluate the path-ordered product,
\begin{equation}
 T_{A+B}(\sigma_f,\sigma_i) \equiv
  P e^{\int_{\sigma_i}^{\sigma_f}d\sigma \left(
 A(\sigma) + B(\sigma)
					\right)},
 \label{A+B}
\end{equation}
with $M\!\times\!M$ matrices $A$ and $B$.
{}From the definition of the path-ordered product, (\ref{A+B}) can be
written as
\begin{align}
  T_{A+B}(\sigma_f,\sigma_i) =
 1+\sum_{n=1}^\infty \int d\sigma_1\cdots d\sigma_n
 &\Bigl(A(\sigma_1)+B(\sigma_1)\Bigr)
 \cdots
 \Bigl(A(\sigma_n)+B(\sigma_n)\Bigr) \nn \\
 &\times\theta(\sigma_f-\sigma_n)\theta(\sigma_n-\sigma_{n-1})
 \cdots\theta(\sigma_{1}-\sigma_i), 
\label{definition of path-ordered product}
\end{align}
where $\theta(\sigma-\sigma')$ is the Heaviside step function. 
Here we take a resummation in 
(\ref{definition of path-ordered product})
by gathering terms containing the same number of $B$'s. 
Clearly, the terms with no $B$ become 
\begin{equation}
   Pe^{\int_{\sigma_i}^{\sigma_f}d\sigma A(\sigma)}
    \Bigl(=T_A(\sigma_f,\sigma_i) \Bigr).
\end{equation}
Similarly, the summation of the terms with a single $B$ becomes 
\begin{equation}
 \int d\sigma_1 
\left(
Pe^{\int_{\sigma_1}^{\sigma_f}d\sigma' A(\sigma')}
\right)
B(\sigma_1)
\left(
Pe^{\int_{\sigma_i}^{\sigma_1}d\sigma' A(\sigma')}
\right). 
\end{equation}
In general, we can easily see that the summation of the terms 
with $n$ $B$'s becomes 
\begin{equation}
 \int d\sigma_1\cdots d\sigma_n 
T_A(\sigma_f,\sigma_n)B(\sigma_n)T_A(\sigma_n,\sigma_{n-1})
B(\sigma_{n-1})\cdots B(\sigma_1)T_A(\sigma_1,\sigma_i).
\end{equation}
Then we can rewrite (\ref{A+B}) as 
\begin{align}
 T_{A+B}(\sigma_f,\sigma_i) &= 
 P_A e^{\int_{\sigma_i}^{\sigma_f}d\sigma B(\sigma)} \nn \\
&\equiv Pe^{\int_{\sigma_i}^{\sigma_f}d\sigma A(\sigma)} 
 + \sum_{n=1}^\infty 
\int d\sigma_1\cdots d\sigma_n 
T_A(\sigma_f,\sigma_n)B(\sigma_n)T_A(\sigma_n,\sigma_{n-1}) \nn \\ 
&\hspace{7cm}
 \times
 B(\sigma_{n-1})\cdots B(\sigma_1)T_A(\sigma_1,\sigma_i), 
\end{align}
where we the symbol $P_A$ expresses 
a kind of path-ordered product 
with replacing the usual Heaviside step function, 
$\theta(\sigma-\sigma')$, by 
the ``transfer matrix'' by $A$, $T_A(\sigma-\sigma')$.

\bibliographystyle{JHEP}
\bibliography{refs}

\end{document}